# A Density Functional Study of the Structures and Energetics of Small Hetero-Atomic Silicon-Carbon Nanoclusters

Prachi Pradhan and Asok K. Ray*
Department of Physics
The University of Texas at Arlington
P.O Box 19059
Arlington, Texas 76019

*email:akr@uta.edu

**Abstract**

The theoretical formalism of local density approximation (LDA) to density functional theory (DFT) has been used to study the electronic and geometric structures of $Si_mC_n$ ($1 \leq m, n \leq 4$) clusters. An all electron 6-311++G** basis set has been used and complete geometry optimizations of different possible structures for a specific cluster have been carried out. Comparisons of the structures and the stabilities of the clusters and their dependence on cluster sizes and stoichiometry have been performed in detail. Binding energies, fragmentation energies, vibrational frequencies, HOMO-LUMO gaps, vertical ionization potentials and vertical electron affinities of the optimized clusters have been investigated and reported in detail. Clusters with equal numbers of silicon and carbon atoms are found to be particularly stable. In particular, based on the simultaneous criteria of high binding energy, high band gap, high ionization potential, and low electron affinity, we believe that $Si_3C_3$ is a candidate for a highly stable or a so-called "magic cluster". Results have been compared with other experimental and theoretical results available in the literature.

## 1. Introduction

Clusters are distinctly different from their bulk state and exhibit many specific properties, which distinguishes their studies as a completely different branch of science named "Cluster Science". Large surface to volume ratio and quantum effects resulting from small dimensions are usually prominent in clusters and ideas like 'super-atoms', 'magic numbers' or 'fission' in clusters have prompted a wide class of scientists to study this 'relatively' new area of the physical sciences. Advances in information technology have enabled scientists to study increasingly complex systems in this area [1-9]. Growing interests in the stabilities of small clusters and the evolution of bulk properties from cluster properties are also due to the emergence of new areas of research called nanoscience and nanotechnology and the resulting potentials in industrial applications, *e.g.* for electronic devices, for data storage, or for fostering chemical reactions [10-13]. Among various types of clusters, simple metal and semiconductor clusters remain to be the most important and widely studied clusters, both experimentally and theoretically. In the area of semiconductor clusters, carbon and silicon, though belonging to the same column of the periodic table, vary significantly in their basic chemical and physical properties. A combination of these two materials silicon-carbide (SiC) or carborundum exists in many different polytypes and, in crystalline phase is a very interesting and important technological material. Silicon carbide (SiC) possesses many favorable properties making it interesting for high-temperature, high-frequency, and high voltage integrated electronics. More specifically, these properties are: wide bandgap, high thermal conductivity (better than for example copper at room temperature), high breakdown electric field strength (approximately 10 times that of Si), high saturation drift velocity (higher than GaAs), high thermal stability and chemical inertness. However, clusters exhibit bonding and behavior, which is far different from that in bulk or molecules.

Our interest in silicon-carbon clusters stems from three points of view. First, in our recent studies, we have observed that carbon clusters trapped inside medium sized silicon cages give rise to highly symmetric, stable cage clusters [14-15]. This work is a part of our continuing investigation in the electronic and geometric structure properties of silicon-carbon clusters to understand the basic principles behind the high stability of such clusters. Second, Si-C clusters have potential usage in the areas of nanoscience and nanotechnology for the development of new materials for the nano-electronics industry. Finally, from a more fundamental perspective, it would be interesting to understand the nature of these hetero-atomic clusters, observe the similarities as well as the differences in silicon and carbon bonding and energetics. We might encounter some novel bonding arrangements, not found either in bulk crystals or stable molecules, yielding lowest energy isomers for SiC clusters. For instance, bulk SiC, has a tetragonal bonding of carbon atom with four nearest silicon neighbors. The bond distances for Si-Si and Si-C are 3.08 Å and 1.89 Å respectively. In all probability, this type of bonding will not be observed for SiC clusters. Silicon is known to prefer multidimensional single bonds and carbon can form single, double and triple bonds. Our objective in this work is also to determine how the structures, bonding and relative stabilities of the clusters are affected by the cluster sizes and the stoichiometry. As the discussions below will show, there have been several works in the literature on small $Si_mC_n$ clusters. *Ours is the first attempt to study these clusters from a completely ab initio point of view with a large basis set* *with added diffuse functions. We also present detailed results on cohesive energies and fragmentation energies and report for the first time results on vertical ionization potentials and vertical electron affinities for $Si_mC_n$ small clusters. Of course, wherever possible, we compare our results with results available in the literature.* We further investigate whether there are certain "magic" compositions or "magic numbers" for which a particular cluster is very stable. First principles simulations, based on density functional theory (DFT) and particularly the local density approximation (LDA), have proved to be a reliable and computationally tractable tool in quantum chemistry and condensed matter physics [16-18]. This formalism is used here to investigate mixed silicon-carbon clusters $Si_mC_n$ ($1 \leq m, n \leq 4$).

## 2. Computational details

One of the primary considerations involved in these calculations is the determination of the methodology, specifically the form of the exchange-correlation potential and the type of basis set to be used. Since experimental results are available for the Si and C atoms [19], we tested our theoretical results (table 1) using various forms of exchange-correlation functionals, such as PW91 [20], B3LYP [21], and SVWN [22], using an all-electron 6-311++G** basis set [23] in the Gaussian '03 suite of programs [24]. Best results were obtained with SVWN (Table 1). Bertolus *et al.* [25] have also concluded, after detailed testing of various functionals with a 6-31G* basis set, that the LSDA functional SVWN is one of the most precise functional for $Si_mC_n$ ($3 \leq m+n \leq 6$) systems. Thus, we have opted to use this functional for all $Si_mC_n$ clusters with an all electron 6-311++G** basis set in the local density approximation (LDA) to density functional theory (DFT). The binding energy per atom for the neutral clusters is calculated from

$$E_b = \{[m\,E\,(Si) + n\,E\,(C)] - [E\,(Si_m C_n)]\} / (m+n) \quad (1)$$

where $E\,(Si_m C_n)$ is the optimized total energy of the cluster.

The fragmentation energy of the clusters into different binary channels is calculated from

$$E_{n \rightarrow (n-m) + m} = E_{n-m} + E_m - E_n \quad , n > m \geq 1 \quad (2)$$

The vertical ionization potential is calculated from

$$VIP = E_{m+n}^{+} - E_{m+n} \quad (3)$$

where $E_{m+n}^{+}$ is the total energy of the corresponding cationic cluster at the neutral geometry. The vertical electron affinity is calculated from

$$VEA = E_{m+n} - E_{m+n}^{-} \quad (4)$$

where $E_{m+n}^{-}$ is the total energy of the corresponding anionic cluster at the neutral geometry. The HOMO-LUMO gap is computed as the energy difference between the highest occupied molecular orbital and the lowest unoccupied molecular orbital. As indicated before, all computations have been performed using the *Gaussian '03* suite of programs [24] using the supercomputing facilities at the University of Texas at Arlington.

**3. Results**

In figures 1-16, we present the structures of lowest energy isomers of the $Si_mC_n$ clusters with ($1 \leq m, n \leq 4$), with the bond lengths in Angstroms. The structures are denoted by (m, n, i), where i runs from 1 to the number of isomers of a particular cluster in decreasing order of stability. Thus, for example, the most stable $Si_4C_4$ cluster is indicated in figure 16 by 4.4.1. All structures are Berny geometry and spin-optimized [26]. Table 2 shows reference values of bond lengths in Ǻ for conventional single and multiple bonds, taken from molecules where H atoms saturate all dangling bonds. Tables 3-18 give the electronic state and binding energy per atom, HOMO-LUMO gap, vertical ionization potential, and vertical electron affinity (all in eV) for each structure. The fragmentation energies along with their different dissociation channels for the most stable clusters are given in table 19. The coordination number for the most stable cluster in each set is shown in table 20. Table 21 gives theoretical and experimental (where available) vibrational frequencies in $cm^{-1}$. We also compare our frequencies with the results obtained by Bertolus *et al.* [25]. Table 22 shows the Mulliken charge distribution analysis [27]. In the following sections, we discuss the results in detail.

### 3.1 SiC dimer

One of the earlier theoretical calculations was by Bauschlicher and Langhoff [28]. They performed configuration interaction calculations using a large ANO [5s4p3d2f1g] basis set and the best estimates for the spectroscopic parameters for the ground $^3\Pi$ state of SiC are $r_e$ = 1.719Å, $\omega_e$ = 962 $cm^{-1}$, and $D_e$ = 4.4 eV. This was followed by an experimental detection of the SiC molecule by Bernath *et al.* [29], using high-resolution Fourier-transform emission spectroscopy techniques. The electronic state was found to be $^3\Pi$, with an equilibrium bond length of 1.719 Å. There have also been other several *ab initio* studies. Martin *et al.* [30] calculated three lowest lying states of SiC, using Hartree-Fock based many-body perturbation theory methods, augmented coupled cluster methods, and configuration interaction methods with different basis sets. The ground state was found to be $^3\Pi$, but the spectroscopic constants were found to be quite sensitive to the electron correlation method employed. Hunsicker and Jones [31], using a combination of density functional and molecular dynamics methods and periodic boundary conditions, also found $^3\Pi$ to be the ground state with a bond length of 1.719 Å. In our calculations, we found the SiC dimer (figure 1) to have a bond length of 1.717 Å with the $^3\Pi$ as the electronic state. Vibrational analysis (table 21) shows a frequency of 986.12 $cm^{-1}$, close to the experimental value 962 $cm^{-1}$. Our dissociation energy for the SiC dimer is 5.505 eV and we note that local density approximation or in general, density functional theory tends to overestimate the cohesive energy.

**3.2 $SiC_2$ isomers**

There have been several studies on this isomer. Robles and Mayorga [32], using 6-31G*//3-21G* basis sets, found $SiC_2$ to be a triangular structure with an apex angle of 67.6° at the Hartree-Fock (HF) level of theory. Hartree-Fock calculations by Grev and Schaefer [33] predicted linear ground state geometry but upon inclusion of correlation effects predicted a triangular ($C_{2v}$) geometry with a total energy of about 1 Kcal lower than the linear system. Exhaustive follow-up studies with very large basis sets and theoretical treatments post CCSD (T) levels confirm that the $^1A_1$ global minimum of $SiC_2$ is a triangle-shaped $C_{2v}$ structure followed by a linear structure. Other theoretical studies of post-HF calculations at MBPT2 and coupled cluster [34] and MBPT4 [35] levels of theory, predict the $C_{2v}$ configuration as the lowest energy state. Sadlej *et. al.* [36] showed the dependency of the basis set and choice of the level of theory on the prediction of true ground state for this cluster. Experimental study by Michalopoulous [37] based on the rotational analysis of resonant two-photon ionization data, concluded the ground state structure to be triangular. Our studies show the $C_{2v}$ triangle-shaped structure with an apical angle of 40.1° to be the ground state and with a binding energy per atom about 0.061 eV lower than the Si-terminated linear geometry. The C-C bond length is 1.262 Å suggesting a probable double bond. As a comparison, the study in Ref. [33] indicates an apex angle of 40.4° with a C-C bond length of 1.28 A. This is followed by a linear structure, with Si atom in middle and with a binding energy per atom of 2.052eV lower than the ground state. The bond lengths for the isomers are between 1.698 Å -1.837 Å for Si-C linkage and between 1.262 Å and 1.282 Å for C-C bond and are comparable with the other theoretical and experimental results.

**3.3 $SiC_3$ isomers**

Theoretical calculations by Robles and Mayorga [32] show $SiC_3$ ground state to be rhomboidal in shape. Experimental Fourier transform microwave (FTM) spectroscopy studies by McCarthy [38] shows $SiC_3$ ground state structure to be also a planar rhomboid as well as a second low-lying cyclic isomer about 5 kcal above the ground state. Alberts *et. al* [39] performed configuration interaction (CI) based studies using a triple-zeta plus double-polarization function (TZ2P) basis set. They also obtained a four-member ring with trans-annular C-C bond of 1.469 Å to be the ground state for $SiC_3$ cluster. Hunsicker and Jones [31], using a combination of density functional and molecular dynamics methods and periodic boundary conditions, also found the rhomboidal ground state. We studied *six* different isomers for this particular cluster. Our studies indicate that Si-terminated chain structure (figure 1.3.1) is the ground state. Given the disagreement, we also carried out HF and post-HF calculations for this particular cluster and found results in agreement with others. But, both LDA and GGA to DFT produce the same linear ground state. We also find large charge transfer occurring between the two carbon atoms separated by 1.305 Å, adjacent to the Si atom, resulting in a strong Coulomb force and hence a very stable linear configuration. Obviously, further experimental and theoretical results are necessary to determine the nature of this cluster. The next structure is rhomboidal, with a trans-annular bond angle of 151.3$^{o}$ and a binding energy per atom 0.072eV lower than the ground state binding energy. This is followed by another rhomboidal structure with a trans-annular bond angle of 65.7$^{o}$ and a binding energy per atom 0.161eV lower than the ground state binding energy. Three other possible structures are also reported in fig. 3 and table 5.

**3.4 $SiC_4$ isomers**

Gordon *et al.* [40], using a combination of rotational spectroscopy of singly substituted isotopic species and large scale coupled-cluster *ab initio* calculations, found $SiC_4$ ground state to be a Si-terminated linear chain. Results by Robles and Mayorga [32] also predict the same. We studied *six* linear, planar and 3D geometries for this cluster. A Si-terminated linear chain (figure 1.4.1) with Si-C linkage of 1.699 Å and C-C bond lengths between 1.271 Å -1.295 Å is the ground state structure. The binding energy per atom is 6.148 eV and is the highest among all the clusters studied in this work. Next, is a fan-shaped planar structure (figure 1.4.2), derived essentially from $SiC_3$ isomer (figure 1.3.2). The binding energy per atom increases from 5.398 eV (for $SiC_3$) to 5.972 eV ($SiC_4$) by the addition of an extra carbon atom. This structure has stretched Si-C bonds of 2.081 Å and C-C bond length between 1.278 Å - 1.372 Å. There are few other low lying states, but mostly planar or linear, showing the preference of C-rich clusters for linear geometry. All these structures have singlet states. Frequency analysis (table 21) indicates that our vibrational frequencies are in good agreement with experimental data as well as other theoretical results [25, 40].

**3.5 $Si_2C$ isomers**

Dierckson *et al*. [41], using Si (6s, 5p) and C (6s, 4p) basis sets augmented with d-polarization functions, and Robles and Mayorga [32], using 6-31G*//3-21G* basis sets, found $Si_2C$ to be linear at the Hartree-Fock (HF) level of theory. However, Dierckson *et al.*'s many body perturbation theory (MBPT) calculations, including all single, double, triple, and quadruple substitutions found a bent structure to be the minimum energy structure at a Si-C-Si angle of 118°, suggesting the importance of electron correlations in such systems. The SiC bond distance was 1.715 Å and the ionization potential and electron affinity were 9.29 eV and – 0.37 eV, respectively. On the other hand, Grev and Schaefer [42], and Largo-Cabrerizo and Flores [43] have shown that HF theory also gives the bent structure to be the minimum by using a double zeta and two sets of polarization functions on each atom. Rittby [44] investigated this system using various basis sets at HF level as well as second order many-body perturbation theory (MBPT2) level and found the ground state to be indeed a closed shell $C_{2v}$ symmetry structure. The bond distance at the MBPT2 level with a 6-311G (2d) basis set was found to be 1.703 Å, with a bond angle of 119.5°. Infrared matrix isolation spectrum of the $Si_2C$ molecule by Kafafi *et al.* [45] concluded that the molecule has $C_{2v}$ symmetry with a lower limit of 110° for the Si-C-Si bond angle. We have identified *three* optimized structures for $Si_2C$ trimer. The lowest energy isomer is a bent singlet $^1A_1$ state (figure 2.1.1) with an apex angle of 136.52°. The general structure agrees also with the results of Bertolus *et al.* [25]. Our Si-C bond lengths are 1.694 Å, slightly lower than the SiC dimer bond length. But, the Si-Si bond length is 3.147 Å, a rather high value. The corresponding distances obtained by Bertolus *et al.* [25], using a smaller basis set, are 1.699 Å and 3.021 Å. Next, a linear structure (figure 2.1.2) with the carbon atom in the middle has a binding energy per atom only 0.002eV lower than the corresponding energy of the bent ground state. The Si-C bond length is comparable with the global minima structure. At the LDA level of theory, these two structures are *basically* degenerate. It is possible that these two isomers would likely coexist in an experimental growth environment, transforming into each other given a small amount of energy. Another carbon terminated asymmetric linear structure has a binding energy per atom of 1.367 eV lower than the corresponding energy of the global minimum. Our theoretical frequencies do not agree well with experimental values and further experimental results will be welcome.

### 3.6 $Si_2C_2$ isomers

Trucks and Bartlett [46] studied the low-lying electronic states of the $Si_2C_2$ system using fourth-order many body perturbation theory (MBPT4). They found the lowest lying structure to be a rhombus, 12 kcal/mol lower than the linear structure. This was followed by another MBPT study by Fitzgerald and Bartlett [47]. In this study, the rhombus was again found to be the ground state structure but a distorted trapezoid was found to be the second most stable structure, the difference in energies being only 4.0 kcal/mol. Second order many body perturbation theory (MP2) studies by Lamertsma *et al*. [48] and MD calculations by Hunsicker and Jones [31] also found the ordering for stability from the rhombic structure followed by the trapezoid and then the linear structure. The same conclusions were reached by Presilla-Marquez *et al*. [49] by MP2 calculations with a 6-311G** basis set. Using multi-configurational-self-consistent-field (MCSCF) wave functions, Robles and Mayorga [32] and Rintelman and Gordon [50] also found the global minimum structure of $Si_2C_2$ to be a rhombus. In our calculations, we optimized *ten* different structures. The lowest energy isomer is a cyclic rhombus with a $^1A_g$ state (figure 2.2.1). The C-C bond length of 1.422 Å is intermediate between a single bond (1.54 Å) and a double bond (1.35 Å). The Si-C bond length of 1.834 Å is similar to a single bond length found in methysilane [51] and comparable to the lengths found in $SiC_2$ [33]. Thus, our results suggest single bonds between silicon and carbon and a probable weak bond between carbons. The optimization process also resulted in four linear chain structures, the linear chain with two carbon atoms in the middle (figure 2.2.2) being the most stable among all chains and about 0.067 eV/atom binding energy lower than the corresponding energy of the ground state. The $^3\Sigma_g$ state also has a strong C-C bonding, with the C-C bond of length 1.276 Å being between the typical double bond (1.35 Å in ethylene) and triple bond (1.21 Å in acetylene) .The Si-C bond of 1.737 Å is close to the double bond value obtained by Schaefer [52]. In general, we find that silicon-terminated chains are more stable than the carbon terminated clusters. A planar distorted trapezoidal structure (figure 2.2.3) is the third most stable structure, with a 0.083eV/atom lower binding energy than the corresponding energy of the rhombic global minima. The C-C bond length is 1.316 Å. Another trapezoidal structure (figure 2.2.5) with Si-C bond of 2.263 Å and C-C bond of 1.272 Å was also found to be one of the ten isomers. Another rhombic structure (figure 2.2.8) with Si-Si linkage of 2.335 Å, indicating single bonding was also observed. We note that, in general, the $Si_2C_2$ cluster, with equal numbers of silicon and carbon atoms, favors planar geometry. The experimental vibrational spectra by Presilla-Marquez *et al*. [49] suggested frequencies of 982.9cm$^{-1}$and 382.2cm$^{-1}$ for the fundamental modes of rhombic $Si_2C_2$. Our calculated frequencies (table 21) compare well with the frequencies obtained by Bertolus *et al.* [25] and the experimental frequencies for the ground state structure.

### 3.7 $Si_2C_3$ isomers

There have been several studies on this cluster. Presilla-Márquez [53] performed Fourier transform infrared measurements in conjunction with *ab initio* calculations at MBPT2/DZP level and reported a linear Si- terminated structure to be the most stable structure. Rittby [54] investigated nine different isomeric structures of $Si_2C_3$ using Hartree-Fock (HF) and second-order many body perturbation (MBPT2) theories. Both the theories predicted the linear ground state. They observed the experimental frequency somewhat larger than the predicted MBPT2/DZP value. Duan *et al.* [55] reported photoelectron spectrum of $Si_2C_3$ cluster along with theoretical studies at Hartree-Fock, hybrid DFT (B3LYP), and multi-configurational-self-consistent-field (MCSCF) levels employing cc-pVDZ basis set. Six linear and eight non-linear isomers were studied and a centro-symmetric linear ground state was observed. The B3LYP and MCSCF gave comparable electron structures, but the HF method predicted significantly different electronic structure. We studied *twelve* different isomers and obtained a different energetic ordering; we conclude in agreement that, the most stable isomer (figure 2.3.1) is the Si-terminated linear chain in $^1\Sigma_g$ state as the ground state structure. The Si-C linkage is 1.691 Å, a slight reduction from 1.699 Å in case of $SiC_4$ (figure 1.4.1). The C-C bonds are 1.286 Å, in the same range as the as found in $SiC_4$ cluster (figure 1.4.1). Comparison with the other five-member clusters, $C_5$ and $SiC_4$ shows that all these structures have linear ground states. They all show distinct features of a linear $C_3$ sub-molecule exhibiting double bonds. Another planar structure with a $C_3$ sub-molecule (figure 2.3.2) is found to have a binding energy per atom 0.425eV less than the corresponding energy of the global minima. It is important to note that the attaching a carbon atom to the $SiC_3$ ground state (5.470 eV), led to a substantial lowering of energy and the linear ground state for $SiC_4$ cluster (6.148 eV). Addition of an extra silicon atom, led to a low lying $Si_2C_3$ isomer (figure 2.3.8) which is about 0.703 eV less that the corresponding global minima. The symmetrical linear chains are more stable than the asymmetric ones. The stability of most of the isomers is due to the formation of multiple strong C-C bonds, which takes precedence over the formation of Si-C or Si-Si bonds. This set of clusters prefers linear geometry and termination by silicon atoms. This also indicates sp hybridization. In table 21, we have compared our vibrational frequencies with the theoretical and experimental data. The predicted values of 907.3 $cm^{-1}$ and 2062.7 $cm^{-1}$, are close to the experimental values of 898.9 $cm^{-1}$ and 1955.2 $cm^{-1}$.

**3.8 $Si_2C_4$ isomers**

Results on five different structures by Froundakis *et al.* [56] and by Hunsicker and Jones [31] indicate a Si-terminated linear chain in $^3\Sigma_g$ electronic state as the ground state structure. Bertolus *et al.* [25] also concluded a linear chain to be the most stable structure. We investigated these as well as several other structures, a total of *twenty-one* isomers in various configurations and found a different energetic ordering as compared to the previously reported results for this cluster. We do find that a linear chain (figure 2.4.1) with terminal silicon atoms in $^3\Sigma_g$ electronic state is the most stable structure. The terminal silicon atoms along with three C-C bonds (1.274 Å, 1.291 Å, and 1.274 Å) in linear arrangement contribute to the stability of this cluster. Next in the energetic ordering, is a t-shape structure (figure 2.4.2) with three C-C bonds (1.282 Å, 1.288 Å, 1.282 Å), and binding energy per atom 0.248 eV less than the stable minima, followed by a distorted hexagonal ring (figure 2.4.3), derived essentially from $C_6$ ground state with binding energy per atom 0.295 eV less than the ground minima. A planar structure (figure 2.4.4) is also derived from a low-lying isomer of the six-atom cluster $C_6$. The stability of this cluster can be attributed to the formation of a rhombus of four carbon atoms with bond lengths (1.410 Å, 1.529 Å). There are other structures mostly common in carbon-rich clusters showing a clear preference of this stoichiometry for linear or planar geometry. Frequency analysis (table 21) indicates that our vibrational frequencies are in good agreement with experimental data as well as theoretical results published in the literature.

### 3.9 $Si_3C$ isomers

Rittby [57] investigated six different isomeric structures of $Si_3C$ using Hartree-Fock (HF) and second-order many body perturbation theories. The ground state structure was a rhomboidal $C_{2v}$ structure with two equivalent silicon atoms and a trans-annular Si-C bond. The same structure was also the ground state structure obtained by Hunsicker and Jones in their density functional calculations with simulated annealing [31], by Bertolus *et al.* [25], and by Robles and Mayorga [32]. We have spin-optimized *seven* different structures and conclude that the most stable isomer (figure 3.1.1) is the fan-shaped rhomboidal structure in $^1A_1$ state with a $Si_2C$ sub molecule (3.1.1) as the ground state structure. The Si-C linkage is 1.760 Å and Si-Si bond is 2.416 Å. Comparison with the other four-member clusters, $C_4$, $SiC_3$, $Si_2C_2$ and $Si_4$ shows that all these structures expect $C_4$ have rhombic or rhomboidal ground states. Another T-shaped structure with $Si_2C$ sub molecule and carbon on top (figure 3.1.2) is found to have a binding energy per atom 0.265 eV less than the corresponding energy of the global minima. Two structures; a rhomboidal $C_{2v}$ structure (figure 3.1.3) with a $Si_3$ silicon sub-molecule and a tetrahedron with a carbon atom on top (figure 3.1.4) are energetically very close to each other and differ by only 0.011 eV in binding energy per atom. A linear chain with one carbon in middle (figure 3.1.5) is the next local minimum with a binding energy per atom 0.592 eV less than the corresponding energy of the global minimum followed by a T-shape structure (figure 3.1.6) with a $Si_3$ sub-molecule. A carbon-terminated chain cluster (figure 3.1.7) is the least stable structure among all the clusters considered in this group. In general, we conclude that a cluster with more Si-C linkage or a cluster with $Si_2C$ sub-molecule is more energetically preferred than the one with $Si_3$ sub-molecule. Also, silicon-terminated chains are energetically preferred over carbon-terminated clusters. The bonding preference thus appears to be in the order Si-C and Si-Si. Frequency analysis (table 21) indicates that our vibrational frequencies are in good agreement with experimental data as well as theoretical results.

### 3.10 $Si_3C_2$ isomers

Froudakis *et al*. [58] studied five different isomers at the MP2 level of theory. This study and the MD-DF calculations of Hunsicker and Jones [31] indicate the lowest lying structure to be a planar pentagon. Bertolus *et al*. [25] and Robles and Mayorga [32] also concluded a planar pentagon to be the most stable structure. We spin-optimized a total of *ten* isomers in different possible multiplicities. Initial geometries were chosen on the basis of $Si_5$ and $C_5$ structures. We conclude that a planar pentagon with a $C_2$ sub molecule (figure 3.2.1) is the ground state structure showing strong multi-center bonding. The strong Si-C bonds of lengths 1.727 and 1.909 Å, along with a stronger C-C bond of length 1.362 Å contribute towards the stability of this cluster. Our vibrational frequencies (table 20) agree well with the experimental and theoretical frequencies reported in the literature. The next stable structure (figure 3.2.2) is a silicon-terminated alternate chain structure with a binding energy per atom 0.364 eV lower than the corresponding energy of the global minimum structure. The two next stable structures are three-dimensional structures; a prism (figure 3.2.3) with C-Si-C angle of 126.99 degrees and a bi-pyramid (figure 3.2.4.) with C-Si-C angle of 88.83 degrees. These structures are energetically close to each other and differ in binding energy per atom by 0.092 eV. The prism is more stable because of the open geometry of the structure. Basically in the prism, carbons cannot form a bond, but overall, this structure is held together due to the presence of several Si-C bonds. A planar dumbbell structure (figure 3.2.6) and a three-dimensional structure (figure 3.2.7) have the same energy and have binding energies per atom 0.621 eV lower than the corresponding energy of the global minimum. The planar structure has a $SiC_2$ sub-molecule with a C-C bond of 1.262 Å but the 3D structure with $C_1$ symmetry also exhibits a C-C bond of 1.354 Å with a shorter Si-C bond of 1.806 Å as compared to other structures and thus is not as stable as some of the other structures. Other less favored structures are also shown in Figure 10. We note again that the stability of a cluster depends on the formation of strong C-C and Si-C bonds, which are more energetically favorable than the Si-Si bonds. For this cluster, planar structures are preferred over three-dimensional structures.

### 3.11 $Si_3C_3$ isomers

Muhlhauser *et al.* [59] investigated up to seventeen different isomers of this clusters by re-optimizing most stable HF geometries at MP2 (TZP) and MP2 (TZ2P) levels. Hunsicker and Jones [31] performed MD-DF calculations on neutral and anionic eight stable geometries. Both studies indicate that the ground structure is a pyramid-like structure with $C_s$ symmetry. We studied *six* isomers and conclude, in agreement with others, that the distorted pentagonal pyramid with a $C_3$ sub-molecule (figure 3.3.1) is the most stable structure. There is a formation of a C-C double bond of length 1.325 Å. The distance between the two base silicon atoms is 3.67 Å, too large for any silicon single bonds. But the two capping atoms have Si-Si bond distance of 2.538 Å well within the range of Si-Si single bond length. The Si-C bonds are 1.858 Å, lie within the range for Si-C single bond length. The stability of this structure is due to high Si-coordination. The vibrational frequencies are given in table 20. There are only two experimental frequencies available and one of our frequencies is close to the lower frequency. Three other three-dimensional structures, close in binding energy per atom to the ground state structure, were also found. One of these structures (figure 3.3.5) showed formation of two C-C bonds of lengths 1.307 Å and 1.358 Å and is the least stable among all the 3D structures. Thus, $Si_3C_3$ with equal numbers of silicon and carbon atoms, prefer overall three-dimensional geometries compared to the planar or linear structures. Stability in this cluster is attributed to the formation of multiple bonds.

### 3.12 $Si_3C_4$ isomers

As far as we know, these clusters have not yet been studied in detail yet. There is only one recent study by Bertolus *et al.* [60], a combination of MD and *ab initio* DFT methods, who identified a pyramid-shaped planar cyclic structure to be the ground state structure. They studied various planar, quasi-planar and 3D structures. We examined a total of *thirteen* structures and obtained a different energetic ordering. However, we also conclude that a slightly distorted pyramid-shaped planar cyclic structure (figure 3.4.1) to be the ground state structure. The C-C bond length is 1.287 Å between the typical double bond (1.35 Å in ethylene) and triple bond (1.21 Å in acetylene) and the Si-Si bond length is 2.251 Å indicating single bonding. Next, is another planar structure (figure 3.4.2) with binding energy per atom 0.033eV less than the ground state minimum. It has four carbon atoms arranged in a planar trapezoidal shape with bond lengths (1.540 Å, 1.560 Å, and 1.366 Å) and capped by silicon atoms. The Si-C linkage is between 1.730 Å -1.890 Å. Next, is a 3D structure (figure 3.4.3) with binding energy per atom 0.215eV less than the minima. The carbon atoms are arranged in planar trapezoidal shape with C-C bond lengths (1.60 Å, 1.450 Å, 1.443 Å), slightly greater than the previous planar arrangement. Following these is another 3D structure (figure 3.4.4) based on the most stable structure found in $Si_3C_3$ cluster, a pyramid-like structure with $C_s$ symmetry. This structure has binding energy per atom 0.226eV less than the minima. Other low-lying structures observed for this cluster (figure 12) are mostly linear or planar. Owing to the large cluster size, many isomers are possible. For the sake of brevity, we have reported here only the most stable structures possible. It is important to note that this cluster has isomers preferring linear or chain geometry. In contrast to other small clusters like $Si_2C_3$ or $Si_2C_4$, the stability no longer is dependent on formation of multiple C-C bonds. It is dependent on the coordination of the silicon atoms in the cluster. As this is a carbon-rich cluster we see mostly $sp^2$ hybridization, but with traces of $sp^3$ hybridization. Our calculated frequencies (table 21) compare well with the frequencies obtained by Bertolus *et al.* [60] for the ground state structure.

### 3.13 $Si_4C$ isomers

The studies so far on this system have been performed, for example, by Bertolus *et al.* [25], Zdetsis *et al.* [61] who identified five structures, and by Hunsicker and Jones [31] who identified two structures for this cluster. We investigated various geometries and report here *seven* most stable isomers. The initial geometries for the isomers were based on a $Si_4$, $Si_5$, and $C_5$ clusters. The most stable structure based on the $Si_5$ ground state is found to be a distorted trigonal bipyramid with a carbon atom at the apex (figure 4.1.1). This is similar to the structure obtained by Robles and Mayorga [32]. The Si-Si linkage is 2.457 Å in between 2.349 Å (at HF level) to 2.606 Å (at MP2 level) as reported by Zdetsis *et al.* [61]. The Si-C bond length is 1.879 Å, slightly longer than the results by Zdetsis *et al.* [61]. Also the bond angle is changed from 109º (at HF level) to 73 º (at MP2 level) to finally 72.4 º (at LDA-DFT level). Also, a planar structure (figure 4.1.2) lies very close, with a binding energy per atom just 0.114eV lower than the corresponding energy of the ground state structure. It is interesting to note that this structure consisting of a $Si_2C$ sub-molecule is more stable than another planar (figure 4.1.5) structure with a $Si_3$ sub-molecule. The formation of two Si-C bonds of length 1.848 Å (figure 4.1.2) contributes to the stability of this structure. The next stable structure is cross-shaped (figure 4.1.3) with four Si-C bonds of 1.821 Å each. A carbon-terminated chain is the least stable structure of all the clusters considered in this category. As expected, these silicon rich clusters prefer three-dimensional bonding as compared to linear or planar structures and a major reason contributing to the stability is the formation of several Si-C bonds followed by Si-Si bonds.

### 3.14 $Si_4C_2$ isomers

Results on four different structures by Froudakis *et al*. [56] and by Hunsicker Jones [31] indicate only one $C_{2v}$ isomer to be a stable minimum. We studied *fourteen* structures including both two-dimensional as well as three-dimensional geometries, expecting the structures to be similar to $Si_6$ and $C_6$ with some strong C-C bonding. We conclude that a 3D distorted pyramid like structure (figure 4.2.1) is the ground state structure, disagreeing with the above two published results in the literature. The Si-C bond length is 1.841 Å and the Si-Si bond lengths are 2.411 Å and 2.086 Å, respectively. The two extreme silicon atoms are separated by 3.25 Å, too long for any bond and the C-C bond length is 1.530 Å. The next stable structure, with a binding energy per atom 0.114 eV lower than the ground state binding energy, is the three-dimensional structure with $C_{2v}$ symmetry quoted by Froudakis [56] as the minimum. The C-C bond length is 1.307 Å and Si-Si bond length is 2.125 Å, with the Si-C bond length being 1.831 Å. The C-C bond of 1.307 Å, between typical double (1.35 Å in ethane) and triple (1.21 Å in ethane) bonds is a major contributing factor towards to the stability of this isomer. Si-Si bond distance is 2.124 Å. This structure is followed by another three-dimensional structure (figure 4.2.3) with $D_{4h}$ symmetry, with a binding energy per atom 0.171 eV lower than the corresponding energy of the ground state. It shows four Si-Si bonds of 2.423 Å. The C-C separation of 1.786 Å and is probably large for any significant bonding. This structure can also be said to have been derived from the $Si_4$ tetrahedron, but with $C_2$ capping. The reason for the stability of this structure is the presence of eight strong Si-C bonds of length 1.932 Å. Several other low-lying structures were also considered as shown in figure 14.

### 3.15 $Si_4C_3$ isomers

As far as we know, these clusters have not yet been studied in detail yet. The only study by Bertolus *et al*. [60] identified a triangular planar pyramid-like structure to be the ground state structure. We identified *ten* isomers for this cluster and conclude also that the triangular planar pyramid-like structure (figure 4.3.1) is in fact the ground state structure. The C-C bond lengths are 1.336 Å (double bond) and 1.544 Å (single bond). The Si-Si linkage is 2.224 Å indicating a single bond formation. Next, is a 3D structure (figure 4.3.2) with binding energy per atom 0.243 eV less than the ground minima. The Si-Si bond length is 2.539 Å and Si-C bond lengths are1.827 Å and 1.803 Å, respectively. Several Si-C bonds and strong C-C bond of 1.27 Å contribute to the stability of this structure. The next structure is also a three-dimensional structure (figure 4.3.3) with a $C_3$ sub molecule at base with C-C bonds of lengths 1.314 Å. Three chain structures were also observed and silicon terminated chains are found to be more stable that carbon terminated chains. One chain structure with three carbon atoms in center (figure 4.3.5) and C-C bonds each of 1.278 Å in length is the most stable among the three linear chains. Other less stable planar structures are also shown in Figure 15. The general tendency of the low-lying $Si_4C_3$ structures tend to be planar. This is in agreement with the results obtained for another seven atomic species; $Si_3C_4$ cluster, wherein we also saw preferences for planar or quasi-planar geometries. This stresses the influence of carbon atoms, as smaller pure Si clusters prefer 3D geometries. Since there is no experimental data available yet on this set of clusters we have compared our calculated frequencies (table 21) with the only other theoretical results available by Bertolus *et al.* [60] for the ground state structure and as indicated, our frequencies compare favorably.

### 3.16 $Si_4C_4$ isomers

These clusters have also not yet been studied in detail yet. The only other study we are aware of is by Bertolus *et al*. [60] who identified a 3D pyramid-like structure, based on $Si_3C_3$ cluster, to be the ground state structure. We have studied various isomers and report here *thirteen* isomers. As previously observed where silicon atoms are equal to number of carbon atoms, symmetry of a particular cluster played an important role in the overall stability. We concentrated on exploring mostly symmetric geometries for this set of clusters. We agree with Bertolus *et al*. [60] that the 3D pyramid-like structure (figure 4.4.1) derived from the $Si_3C_3$ ground state (figure 3.3.1) in $^1A^`$ electronic state is the most stable structure. There is a formation of two C-C bonds, one double bond of 1.27 Å bond length and a partial double bond of 1.404 Å. The SiC linkage is 1.932 Å. The Si-Si distance is 2.45 Å (can form a stretched Si-Si bond) and 3.55 Å (impossible to form any kind of bonding). Next, is a close-lying planar t-shaped structure (figure 4.4.2) with a $Si_3$ sub-molecule with binding energy per atom 0.187 eV less than the ground state structure. The four carbon atoms in a row with a C-C bond length of 1.259 Å each along with silicon termination contribute towards the high stability of this structure. Three three-dimensional structures were also observed out of which 3D trapezoid (figure 4.4.3) was the most stable, with four C-C bonds of 1.466 Å each. These bonds contribute towards the stability of this isomer. A cubic structure (figure 4.4.6) with no C-C bonds was found to be least stable among the three dimensional structures. Several other planar and linear structures were observed as shown in figure 16. Our theoretical frequencies compare favorably with the results of Bertolus *et al.* [60] for the ground state structure.

## 4. Discussions

### 4.1 Bonding and energetics

We now summarize the major trends observed for the lowest energy isomers. Most important trend seen is strong tendency for the C-rich clusters to be chainlike, whereas the Si-rich structures are either planar or 3D. This is universally true, except for the cases of $SiC_2$ and $Si_3C_4$ cluster which are planar, instead of chain-like. The linear structures exhibit multiple bonding which is the dominating factor in the energetics of C-rich clusters. In planar and 3D structures, very little multiple bond character is observed. Instead, these clusters prefer to have multiple single bonds. Mostly, the SiC bonds are favored over the C-C bonds, in a way that the carbon atoms tend to separate more from each other in order to form maximum SiC bonds. In general, for a particular set of clusters, C-C bond lengths in the structures decreases with the increase in carbon atoms. We also calculated the coordination number, $\gamma$, defined as the average number of bonds for each atom in a cluster (table 20). Coordination is high in case of structures forming multiple bonding. For instance, in the case of $Si_3C_3$ and $Si_4C_4$ clusters the coordination number is very high, 3.3 and 3.5 respectively. As expected, linear structures have less coordination as compared to planar and 3D structures.

As far as the binding energy per atom is concerned, for a particular set of clusters with fixed number of silicon atoms, the binding energy per atom increases with the increase in the number of carbon atoms. The clusters with equal number of silicon and carbon atoms are particularly stable. The binding energy per atom increases from 2.752 eV for SiC to 5.09 eV for $Si_2C_2$ to 5.482 eV for $Si_3C_3$ to 5.526 eV for $Si_4C_4$ clusters. Comparing the clusters, three-atoms $SiC_2$ and $Si_2C$, four-atoms $SiC_3$, $Si_2C_2$, and $Si_3C$, five-atoms $SiC_4$ , $Si_2C_3$ , $Si_3C_2$ , and $Si_4C$, six-atoms $Si_2C_4$ , $Si_3C_3$, and $Si_4C_2$, and seven atoms $Si_3C_4$ and $Si_4C_3$, we note that if the carbon atoms are replaced by silicon atoms, the binding energies tend to decrease. This agrees with the fact that the binding energy per bond increases from Si-Si, Si-C to C-C bonds. In figure 17, we have plotted the binding energy per atom versus the number of atoms in the cluster. In cases where the number of atoms is equal in two clusters, we have chosen the cluster with the higher binding energy per atom. As seen from the graph, the binding energy has an oscillatory pattern, similar to metal clusters. A noticeable peak is for the five-atom cluster $SiC_4$. As noted before, this carbon-rich penta-atomic cluster $SiC_4$ is a linear chain with three C-C bonds for the ground state structure and the highest binding energy per atom of 6.148 eV. Next, most stable structure is the hexa-atomic cluster $Si_2C_4$. The ground state structure is again a linear chain structure with three C-C bonds and a binding energy per atom of 5.971 eV. In general all these clusters exhibit high binding energies.

The fragmentation energies of the ground state clusters into different possible binary channels summarized in table 19. In general, the preferred channel of decay is breaking up of a cluster to separate a single silicon atom. The bond dissociation energy (figure 18), defined as the lower of the difference in total energy between $Si_mC_n$, and $Si_{m-1}C_n$ + Si or $Si_mC_{n-1}$ + C is also plotted in figure 18. The graph shows alternating behavior with higher values at odd clusters expect for $Si_3C_4$ cluster. The $SiC_2$ cluster has the highest dissociation energy of 9.920 eV.

**4.2 HOMO-LUMO gap**

The highest occupied molecular orbital - lowest unoccupied molecular orbital (HOMO-LUMO) gap for the neutral clusters are also given in tables 3-18. *No other published data on these gaps exists in the literature apart from the Hartree-Fock results by Robles and Mayorga on some of these clusters* [32]. Their gaps range from about 7 to 10 eV. These gaps are significantly larger than the gaps in our calculations and the differences are believed to be due to the different theoretical formalisms employed, namely Hartree-Fock theory versus density functional theory. The six-atom cluster $Si_3C_3$, has the highest gap of 3.619 eV. The possible cause for such high gap for $Si_3C_3$ structure is the high multi-center partial ionic and covalent bonding exhibited by it. Another six-atom cluster $Si_4C_2$ also has a rather high gap of 3.530 eV. In figure 19, we have plotted the HOMO-LUMO gap versus the number of atoms in the cluster. An odd-even oscillation exists in the variation of the HOMO-LUMO gap with respect to cluster size. The LDA-DFT approach is well known to underestimate the band gaps of a material and these clusters can indeed qualify as the wide band gap semiconductor clusters similar to their bulk counterparts.

**4.3 Vertical ionization potentials and electron affinities**

*No published data, either theoretical or experimental, exists in the literature for these quantities except for SiC dimer and $SiC_2$.* The vertical ionization potentials (VIP) and the vertical electron affinities (VEA) of the clusters are also shown in tables 3-18. Our vertical ionization potential for the SiC dimer, 9.758 eV, is in the experimental range of 8 to 10 eV. Similarly our vertical ionization potential for $SiC_2$ isomer, 10.396 eV, is in experimental range of 8.9 to 10.4 eV. The vertical ionization potentials as a function of cluster size are shown in figure 19. The saw-tooth behavior is indicated by higher IP's for odd number of atoms; $SiC_2$, $SiC_4$ and $Si_3C_4$ clusters. The three-atom cluster $SiC_2$ possesses the highest ionization potential of 10.396 eV.

The vertical electron affinities, as a function of cluster size are shown in figure 19. A saw-tooth behavior is indicated by higher EA's for even number of atoms; SiC, $SiC_3$ and $Si_2C_4$ clusters. Again, we find an alternating behavior and clusters with lower ionization potentials are found to have higher electron affinities. Five atom cluster $SiC_4$ with highest binding energy per atom in the entire set of clusters has the lowest electron affinity of 1.371 eV.

### 4.4 Charge distribution

Table 22 shows the Mulliken charge distributions [27] for the ground state structures of the $Si_mC_n$ neutral clusters. The geometry and stability of the clusters are closely related to the bonding features and charge distributions. The SiC dimer shows than the carbon atom gains negative charge whereas the silicon atom loses charge, as expected from the electronegativity considerations. The $Si_2C_4$ cluster shows significant charge transfer from silicon to carbon atoms. This is true for all clusters except for the $Si_4C_n$ clusters. Here one of the silicon atoms gains slightly negative charge. For example, for $Si_4C$ structure, the bottom silicon atom exactly opposite to the carbon atom carries partial negative charge while the other three silicon atoms in the triangle have equal positive charges. As we expect from carbon's higher electronegativity, the silicon atom carries less negative charge than the carbon atom. For $Si_4C_2$ cluster, the silicon atom forming the $SiC_2$ sub-molecule in the structure gains negative charge. For the $Si_4C_3$ cluster, the silicon atom in the middle of the chain segment in the structure gains slight negative charge. In the case of $Si_4C_4$ cluster also, the silicon atom at the apex of the structure gains some negative charge. Chain structures with alternating silicon and carbon atoms show alternating positive and negative charges.

### 4.5 Stability

Finally, we note that the $Si_3C_3$ cluster has a rather high ionization potential of 9.664 eV and a low electron affinity of 0.918 eV. Based on the simultaneous criteria of high binding energy, high band gap, high ionization potential, low electron affinity, and high symmetry we believe that $Si_3C_3$ is a candidate for a "magic cluster", though the concept of magicity is not well defined for semiconductor clusters.

**5. Conclusions**

In summary, the formalism of LDA-DFT has been used to study neutral hetero-atomic silicon-carbide clusters. In particular, $Si_3C_3$ cluster is a candidate for a highly stable or a "magic" cluster and clusters with equal number of silicon and carbon atoms appear to be particularly stable. We also conclude that stoichiometry and bonding play an important role in the stability of a cluster. The ratio of $sp^2$ to $sp^3$ hybridization, formation of strong C-C bonds, several Si-C bonds and the coordination of atoms directly affects the preference for a particular geometry and in turn the stability of a cluster. The charge distribution analysis shows significant charge transfer from silicon to carbon atoms, rendering a partial ionic character to the systems. Thus, the stability of these clusters can be attributed to mixed covalent and ionic bonding.

Finally, the authors gratefully acknowledge partial support from the Welch Foundation, Houston, Texas (Grant No. Y-1525).

**Table 1.** Ionization potentials and electron affinities of Si and C atoms.

| Method | Atom | IP | EA |
|---|---|---|---|
| LDA/6-311++G** | Si | 8.667 | 2.003 |
| Expt. | Si | 8.151 | 1.385 |
| LDA/6-311++G** | C | 12.222 | 2.204 |
| Expt. | C | 11.260 | 1.262 |

**Table 2.** Reference values of bond lengths (in Å) for conventional single and multiple bonds, taken from molecules where H atoms saturate all dangling bonds.

| Species | Length |
|---|---|
| C-C | 1.54 |
| C=C | 1.35 |
| C≡C | 1.21 |
| Si-Si | 2.34 |
| Si=Si | 2.11 |
| Si-C | 1.83 |
| Si=C | 1.73 |

**Table 3.** Electronic state, binding energy per atom, homo-lumo gap, vertical ionization potential, and vertical electron affinity, (all in eV) for SiC cluster.

| Structure | State | $E_b/n$ | GAP | VIP | VEA |
|---|---|---|---|---|---|
| 1.1.1 | $^3\Pi$ | 2.752 | 0.048 | 9.758 | 3.303 |

**Table 4.** Electronic states, binding energies per atom, homo-lumo gaps, vertical ionization potentials, and vertical electron affinities (all in eV) for $SiC_2$ clusters.

| Structure | State | $E_b/n$ | GAP | VIP | VEA |
|---|---|---|---|---|---|
| 1.2.1 | $^1A_1$ | 5.141 | 2.00 | 10.396 | 1.830 |
| 1.2.2 | $^1\Sigma$ | 5.080 | 1.893 | 10.414 | 0.260 |
| 1.2.3 | $^1\Sigma_g$ | 3.089 | 0.891 | 10.787 | 3.244 |

**Table 5.** Electronic states, binding energies per atom, homo-lumo gaps, vertical ionization potentials, and vertical electron affinities (all in eV) for $SiC_3$ clusters.

| Structure | State | $E_b/n$ | GAP | VIP | VEA |
|---|---|---|---|---|---|
| 1.3.1 | $^3\Sigma$ | 5.470 | 0.810 | 9.966 | 3.339 |
| 1.3.2 | $^1A_1$ | 5.398 | 0.682 | 10.026 | 2.837 |
| 1.3.3 | $^3B_1$ | 5.309 | 1.343 | 8.574 | 3.201 |
| 1.3.4 | $^1A_1$ | 5.225 | 0.331 | 9.458 | 3.255 |
| 1.3.5 | $^3A$ | 5.203 | 0.722 | 9.551 | 2.470 |
| 1.3.6 | $^3\Sigma$ | 4.559 | 0.569 | 10.505 | 3.973 |

**Table 6.** Electronic states, binding energies per atom, homo-lumo gaps, vertical ionization potentials, and vertical electron affinities (all in eV) for $SiC_4$ clusters.

| Structure | State | $E_b/n$ | GAP | VIP | VEA |
|---|---|---|---|---|---|
| 1.4.1 | $^1\Sigma$ | 6.148 | 1.825 | 10.284 | 1.371 |
| 1.4.2 | $^1A_1$ | 5.972 | 2.358 | 10.635 | 2.253 |
| 1.4.3 | $^1A_1$ | 5.741 | 0.920 | 9.862 | 2.778 |
| 1.4.4 | $^1\Sigma_g$ | 5.540 | 2.093 | 10.672 | 3.038 |
| 1.4.5 | $^1A_1$ | 5.377 | 1.118 | 10.833 | 1.919 |
| 1.4.6 | $^1A_1$ | 4.937 | 0.471 | 10.111 | 4.116 |

**Table 7.** Electronic states, binding energies per atom, homo-lumo gaps, vertical ionization potentials, and vertical electron affinities (all in eV) for $Si_2C$ clusters.

| Structure | State | $E_b/n$ | GAP | VIP | VEA |
|---|---|---|---|---|---|
| 2.1.1 | $^1A_1$ | 4.348 | 2.673 | 9.683 | 1.522 |
| 2.1.2 | $^1\Sigma_g$ | 4.346 | 2.573 | 9.624 | 1.620 |
| 2.1.3 | $^1\Sigma$ | 2.981 | 1.1 | 9.418 | 2.297 |

**Table 8.** Electronic states, binding energies per atom, homo-lumo gaps, vertical ionization potentials, and vertical electron affinities (all in eV) for $Si_2C_2$ clusters.

| Structure | State | $E_b/n$ | GAP | VIP | VEA |
|---|---|---|---|---|---|
| 2.2.1 | $^1A_g$ | 5.09 | 1.902 | 9.569 | 2.059 |
| 2.2.2 | $^3\Sigma_g$ | 5.023 | 0.668 | 8.339 | 2.611 |
| 2.2.3 | $^1A`$ | 5.007 | 0.884 | 8.273 | 2.450 |
| 2.2.4 | $^1A_1$ | 4.716 | 0.321 | 8.828 | 3.030 |
| 2.2.5 | $^3A_2$ | 4.631 | 0.578 | 7.933 | 2.019 |
| 2.2.6 | $^3A_2$ | 4.473 | 0.754 | 9.403 | 3.171 |
| 2.2.7 | $^3\Sigma$ | 4.423 | 0.720 | 9.759 | 3.513 |
| 2.2.8 | $^3B_{3U}$ | 4.353 | 1.244 | 9.476 | 2.210 |
| 2.2.9 | $^3\Sigma$ | 4.048 | 1.099 | 9.691 | 3.175 |
| 2.2.10 | $^3\Sigma$ | 3.190 | 0.646 | 9.814 | 3.996 |

**Table 9.** Electronic states, binding energies per atom, homo-lumo gaps, vertical ionization potentials, and vertical electron affinities (all in eV) for $Si_2C_3$ clusters.

| Structure | State | $E_b/n$ | GAP | VIP | VEA |
|---|---|---|---|---|---|
| 2.3.1 | $^1\Sigma_g$ | 5.746 | 2.069 | 9.124 | 2.200 |
| 2.3.2 | $^3A$ | 5.321 | 0.331 | 9.021 | 2.610 |
| 2.3.3 | $^1A_1$ | 5.243 | 2.680 | 11.120 | 2.333 |
| 2.3.4 | $^3B_2$ | 5.213 | 0.325 | 9.346 | 3.716 |
| 2.3.5 | $^1\Sigma$ | 5.158 | 2.110 | 12.493 | 2.441 |
| 2.3.6 | $^3A$ | 5.142 | 0.681 | 9.240 | 2.964 |
| 2.3.7 | $^3B_2$ | 5.095 | 0.306 | 8.492 | 1.599 |
| 2.3.8 | $^1\Sigma$ | 5.043 | 1.226 | 9.486 | 2.430 |
| 2.3.9 | $^1A_1$ | 4.850 | 0.536 | 8.683 | 3.560 |
| 2.3.10 | $^1\Sigma$ | 4.351 | 1.484 | 8.529 | 3.657 |
| 2.3.11 | $^1A_1$ | 4.420 | 1.349 | 8.929 | 3.499 |
| 2.3.12 | $^3B_2$ | 3.906 | 0.085 | 12.992 | 3.815 |
| 2.3.13 | $^1\Sigma_g$ | 3.832 | 1.226 | 9.663 | 3.892 |

**Table 10.** Electronic states, binding energies per atom, homo-lumo gaps, vertical ionization potentials, and vertical electron affinities (all in eV) for $Si_2C_4$ clusters.

| Structure | State | $E_b/n$ | GAP | VIP | VEA |
|---|---|---|---|---|---|
| 2.4.1 | $^3\Sigma_g$ | 5.971 | 0.509 | 8.053 | 2.969 |
| 2.4.2 | $^1A_1$ | 5.723 | 0.649 | 8.778 | 3.574 |
| 2.4.3 | $^3B_1$ | 5.676 | 0.483 | 8.853 | 3.436 |
| 2.4.4 | $^3A_u$ | 5.644 | 0.911 | 8.146 | 2.492 |
| 2.4.5 | $^1A_g$ | 5.627 | 4.350 | 9.787 | 1.607 |
| 2.4.6 | $^3B_{2g}$ | 5.621 | 0.131 | 8.829 | 3.544 |
| 2.4.7 | $^1A_g$ | 5.608 | 1.189 | 10.483 | 2.876 |
| 2.4.8 | $^3\Sigma$ | 5.529 | 0.649 | 9.656 | 3.836 |
| 2.4.9 | $^3A_2$ | 5.521 | 0.302 | 9.072 | 3.490 |
| 2.4.10 | $^3A_2$ | 5.481 | 0.341 | 8.527 | 3.219 |
| 2.4.11 | $^3B_{1g}$ | 5.462 | 0.412 | 7.502 | 2.463 |
| 2.4.12 | $^3\Sigma$ | 5.406 | 0.598 | 9.719 | 3.472 |
| 2.4.13 | $^3\Sigma$ | 4.351 | 0.677 | 9.133 | 3.542 |
| 2.4.14 | $^3A_2$ | 5.057 | 0.394 | 7.341 | 5.450 |
| 2.4.15 | $^3A_2$ | 5.013 | 0.415 | 9.756 | 4.179 |
| 2.4.16 | $^3\Sigma$ | 4.956 | 0.500 | 7.555 | 3.192 |
| 2.4.17 | $^3A_2$ | 4.851 | 0.905 | 10.026 | 3.663 |
| 2.4.18 | $^3\Sigma$ | 4.784 | 0.650 | 9.888 | 3.955 |
| 2.4.19 | $^1A_1$ | 4.622 | 0.157 | 9.298 | 4.255 |
| 2.4.20 | $^3\Sigma$ | 4.574 | 0.822 | 9.509 | 4.216 |
| 2.4.21 | $^3B_{1g}$ | 4.381 | 0.360 | 9.362 | 3.997 |

**Table 11.** Electronic states, binding energies per atom, homo-lumo gaps, vertical ionization potentials, and vertical electron affinities (all in eV) for $Si_3C$ clusters.

| Structure | State | $E_b/n$ | GAP | VIP | VEA |
|---|---|---|---|---|---|
| 3.1.1 | $^1A_1$ | 4.464 | 1.664 | 8.918 | 2.133 |
| 3.1.2 | $^1A_1$ | 4.199 | 0.222 | 7.733 | 2.474 |
| 3.1.3 | $^3B_1$ | 3.942 | 0.441 | 8.297 | 2.507 |
| 3.1.4 | $^3A$ | 3.931 | 0.504 | 7.257 | 3.365 |
| 3.1.5 | $^3\Sigma$ | 3.872 | 0.739 | 9.060 | 2.868 |
| 3.1.6 | $^3B_1$ | 3.277 | 0.291 | 9.032 | 3.674 |
| 3.1.7 | $^3\Sigma$ | 3.045 | 0.724 | 9.499 | 3.545 |

**Table 12.** Electronic states, binding energies per atom, homo-lumo gaps, vertical ionization potentials, and vertical electron affinities (all in eV) for $Si_3C_2$ clusters.

| Structure | State | $E_b/n$ | GAP | VIP | VEA |
|---|---|---|---|---|---|
| 3.2.1 | $^1A_1$ | 5.08 | 1.38 | 8.282 | 1.906 |
| 3.2.2 | $^1A`$ | 4.716 | 2.557 | 11.864 | 3.933 |
| 3.2.3 | $^3B_2$ | 4.603 | 0.378 | 8.647 | 3.707 |
| 3.2.4 | $^3B_1$ | 4.511 | 0.874 | 8.641 | 3.854 |
| 3.2.5 | $^1A_1$ | 4.467 | 0.320 | 9.163 | 3.891 |
| 3.2.6 | $^1A_1$ | 4.459 | 0.169 | 8.914 | 3.572 |
| 3.2.7 | $^1A_1$ | 4.459 | 0.540 | 8.669 | 2.364 |
| 3.2.8 | $^1A_1$ | 4.324 | 1.361 | 9.189 | 3.002 |
| 3.2.9 | $^3B_2$ | 3.981 | 0.485 | 9.019 | 3.359 |
| 3.2.10 | $^1\Sigma$ | 3.213 | 0.993 | 9.514 | 4.225 |

**Table 13.** Electronic states, binding energies per atom, homo-lumo gaps, vertical ionization potentials, and vertical electron affinities (all in eV) for $Si_3C_3$ clusters.

| Structure | State | $E_b/n$ | GAP | VIP | VEA |
|---|---|---|---|---|---|
| 3.3.1 | $^1A$ | 5.482 | 3.619 | 9.664 | 0.918 |
| 3.3.2 | $^1A_1$ | 5.410 | 0.201 | 7.723 | 2.836 |
| 3.3.3 | $^1A$ | 5.349 | 3.887 | 9.185 | 2.191 |
| 3.3.4 | $^3A$ | 5.219 | 0.306 | 8.397 | 2.836 |
| 3.3.5 | $^3A$ | 5.190 | 0.049 | 8.353 | 3.374 |
| 3.3.6 | $^3\Sigma$ | 4.430 | 1.048 | 9.041 | 3.364 |

**Table 14.** Electronic states, binding energies per atom, homo-lumo gaps, vertical ionization potentials, and vertical electron affinities (all in eV) for $Si_3C_4$ clusters.

| Structure | State | $E_b/n$ | GAP | VIP | VEA |
|---|---|---|---|---|---|
| 3.4.1 | $^1A_1$ | 5.745 | 2.354 | 8.941 | 1.924 |
| 3.4.2 | $^1A_1$ | 5.712 | 3.120 | 10.423 | 1.403 |
| 3.4.3 | $^1A_1$ | 5.530 | 0.892 | 7.980 | 2.348 |
| 3.4.4 | $^1A_1$ | 5.519 | 0.347 | 7.350 | 2.114 |
| 3.4.5 | $^1A_1$ | 5.498 | 1.138 | 8.640 | 2.760 |
| 3.4.6 | $^1A_1$ | 5.496 | 0.371 | 8.192 | 3.333 |
| 3.4.7 | $^1\Sigma_g$ | 5.410 | 0.946 | 7.908 | 2.253 |
| 3.4.8 | $^1A^`$ | 5.242 | 0.421 | 8.033 | 2.739 |
| 3.4.9 | $^1A_1$ | 5.185 | 0.899 | 8.491 | 3.283 |
| 3.4.10 | $^1\Sigma$ | 5.117 | 1.104 | 8.727 | 3.395 |
| 3.4.11 | $^1A_2$ | 4.796 | 0.414 | 8.756 | 3.616 |
| 3.4.12 | $^1A_1$ | 4.423 | 0.341 | 8.958 | 4.381 |
| 3.4.13 | $^3\Pi$ | 4.167 | 0.086 | 8.679 | 4.428 |

**Table 15.** Electronic states, binding energies per atom, homo-lumo gaps, vertical ionization potentials, and vertical electron affinities (all in eV) for $Si_4C$ clusters.

| Structure | State | $E_b/n$ | GAP | VIP | VEA |
|---|---|---|---|---|---|
| 4.1.1 | $^1A^`$ | 4.444 | 2.389 | 8.15 | 2.448 |
| 4.1.2 | $^1A_1$ | 4.304 | 1.079 | 8.143 | 2.403 |
| 4.1.3 | $^3B_2$ | 4.232 | 0.632 | 7.448 | 2.194 |
| 4.1.4 | $^1A_1$ | 4.030 | 0.481 | 8.368 | 3.199 |
| 4.1.5 | $^3A$ | 3.567 | 0.277 | 8.778 | 3.605 |
| 4.1.6 | $^1A_1$ | 3.303 | 0.095 | 8.689 | 4.022 |

**Table 16.** Electronic states, binding energies per atom, homo-lumo gaps, vertical ionization potentials, and vertical electron affinities (all in eV) for $Si_4C_2$ clusters.

| Structure | State | $E_b$/n | GAP | VIP | VEA |
|---|---|---|---|---|---|
| 4.2.1 | $^1A$` | 5.270 | 3.530 | 9.795 | 0.337 |
| 4.2.2 | $^1A_1$ | 5.156 | 2.353 | 8.670 | 1.462 |
| 4.2.3 | $^1A_1$ | 5.099 | 3.651 | 9.869 | 0.650 |
| 4.2.4 | $^1A_1$ | 4.886 | 0.979 | 8.185 | 2.230 |
| 4.2.5 | $^1A_1$ | 4.683 | 0.282 | 7.89 | 3.285 |
| 4.2.6 | $^3B_{3g}$ | 4.621 | 0.761 | 7.798 | 3.412 |
| 4.2.7 | $^1A_g$ | 4.522 | 2.320 | 9.090 | 3.478 |
| 4.2.8 | $^3A$`` | 4.520 | 0.183 | 7.509 | 3.534 |
| 4.2.9 | $^3\Sigma_g$ | 4.418 | 0.496 | 8.215 | 2.349 |
| 4.2.10 | $^3A$`` | 4.389 | 0.530 | 8.601 | 3.402 |
| 4.2.11 | $^1A_1$ | 4.308 | 0.522 | 8.664 | 3.695 |
| 4.2.12 | $^3B1_g$ | 4.235 | 0.478 | 7.851 | 3.241 |
| 4.2.13 | $^3A_2$ | 4.090 | 0.417 | 8.395 | 3.686 |
| 4.2.14 | $^1A_g$ | 3.592 | 0.423 | 8.831 | 3.893 |

**Table 17.** Electronic states, binding energies per atom, homo-lumo gaps, vertical ionization potentials, and vertical electron affinities (all in eV) for $Si_4C_3$ clusters.

| Structure | State | $E_b$/n | GAP | VIP | VEA |
|---|---|---|---|---|---|
| 4.3.1 | $^1A$ | 5.353 | 2.256 | 8.415 | 1.893 |
| 4.3.2 | $^3A$ | 5.110 | 0.214 | 8.194 | 3.337 |
| 4.3.3 | $^3A$ | 5.064 | 0.150 | 7.684 | 2.795 |
| 4.3.4 | $^3A$`` | 5.017 | 0.554 | 6.862 | 2.722 |
| 4.3.5 | $^1\Sigma_g$ | 4.859 | 0.684 | 7.920 | 3.435 |
| 4.3.6 | $^1\Sigma_g$ | 4.842 | 2.392 | 8.659 | 2.484 |
| 4.3.7 | $^1A_1$ | 4.638 | 0.227 | 8.773 | 4.204 |
| 4.3.8 | $^3A$` | 4.595 | 0.947 | 7.801 | 4.024 |
| 4.3.9 | $^3A_2$ | 4.367 | 0.372 | 9.147 | 4.159 |
| 4.3.10 | $^3A$ | 4.105 | 0.192 | 7.972 | 3.921 |

**Table 18.** Electronic states, binding energies per atom, homo-lumo gaps, vertical ionization potentials, and vertical electron affinities (all in eV) for $Si_4C_4$ clusters.

| Structure | State | $E_b/n$ | GAP | VIP | VEA |
|---|---|---|---|---|---|
| 4.4.1 | $^1A^`$ | 5.626 | 1.899 | 8.639 | 2.220 |
| 4.4.2 | $^1A_1$ | 5.439 | 0.230 | 8.406 | 3.541 |
| 4.4.3 | $^3A$ | 5.276 | 0.714 | 8.987 | 3.531 |
| 4.4.4 | $^3A$ | 5.230 | 0.50 | 8.867 | 3.637 |
| 4.4.5 | $^1A_1$ | 5.124 | 0.495 | 11.913 | 3.966 |
| 4.4.6 | $^1A$ | 5.079 | 0.836 | 8.953 | 3.346 |
| 4.4.7 | $^3\Sigma_g$ | 5.052 | 0.409 | 7.445 | 3.173 |
| 4.4.8 | $^1A_1$ | 5.022 | 0.255 | 8.264 | 3.883 |
| 4.4.9 | $^1A_g$ | 5.012 | 0.933 | 9.204 | 0.709 |
| 4.4.10 | $^1A_g$ | 4.896 | 0.202 | 7.676 | 3.672 |
| 4.4.11 | $^3B_1$ | 4.893 | 0.118 | 8.014 | 4.272 |
| 4.4.12 | $^3A_2$ | 4.773 | 0.233 | 8.250 | 4.134 |
| 4.4.13 | $^3\Sigma$ | 4.611 | 1.022 | 8.638 | 3.480 |

**Table 19.** Fragmentation energies (in eV) for the ground state $Si_mC_n$ clusters.

| Initial Cluster | Channel | Fragmentation Energy |
|---|---|---|
| SiC | Si + C | 5.505 |
| $SiC_2$ | Si + $C_2$ | 7.679 |
| | SiC + C | 9.920 |
| $SiC_3$ | Si + $C_3$ | 5.134 |
| | SiC + $C_2$ | 8.632 |
| | $SiC_2$ + C | 6.458 |
| $SiC_4$ | Si + $C_4$ | 7.338 |
| | SiC+$C_3$ | 8.491 |
| | $SiC_2$ + $C_2$ | 7.574 |
| | $SiC_3$ + C | 8.862 |
| $Si_2C$ | $Si_2$ + C | 9.032 |
| | Si + SiC | 7.540 |
| $Si_2C_2$ | $Si_2$ + $C_2$ | 8.608 |
| | $Si_2C$ + C | 7.321 |
| | Si + $SiC_2$ | 4.942 |
| $Si_2C_3$ | $Si_2$ + $C_3$ | 7.969 |
| | $Si_2C$ + $C_2$ | 7.940 |
| | $Si_2C_2$+C | 8.364 |
| | Si + $SiC_3$ | 6.848 |
| $Si_2C_4$ | $Si_2$ + $C_4$ | 8.407 |
| | $Si_2C$ + $C_3$ | 6.034 |
| | $Si_2C_2 + C_2$ | 7.715 |
| | $Si_2C_3 + C$ | 7.097 |
| | $Si + SiC_4$ | 5.083 |
| $Si_3C$ | $Si_3 + C$ | 8.841 |
| | $Si_2 + SiC$ | 8.340 |
| | $Si + Si_2C$ | 4.813 |
| $Si_3C_2$ | $Si_3 + C_2$ | 8.639 |
| | $Si_2 + SiC_2$ | 5.965 |
| | $Si + Si_2C_2$ | 5.036 |
| | $Si_3C + C$ | 7.544 |
| $Si_3C_3$ | $Si_3 + C_3$ | 7.128 |
| | $Si_3C + C_2$ | 7.290 |
| | $Si_3C_2 + C$ | 7.492 |
| | $Si_2 + SiC_3$ | 6.999 |
| | $Si + Si_2C_3$ | 4.163 |
| $Si_3C_4$ | $Si + Si_2C_4$ | 4.462 |
| | $Si_2 + SiC_4$ | 5.532 |
| | $Si_3 + C_4$ | 7.865 |
| | $Si_3C + C_3$ | 5.683 |
| | $Si_3C_2+C_2$ | 7.141 |
| | $Si_3C_3+C$ | 7.395 |
| $Si_4C$ | $Si + Si_3C$ | 4.362 |
| | $Si_2 + Si_2C$ | 5.162 |
| | $Si_3 + SiC$ | 7.698 |
| | $Si_4 + C$ | 7.887 |
| $Si_4C_2$ | $Si + Si_3C_2$ | 5.275 |
| | $Si_2 + Si_2C_2$ | 6.298 |
| | $Si_3 + SiC_2$ | 6.235 |
| | $Si_4 + C_2$ | 8.598 |
| | $Si_4C + C$ | 8.456 |
| $Si_4C_3$ | $Si + Si_3C_3$ | 4.580 |
| | $Si_2 + Si_2C_3$ | 4.731 |
| | $Si_3 + SiC_3$ | 6.574 |
| | $Si_4 + C_3$ | 6.392 |
| | $Si_4C + C_2$ | 7.508 |
| | $Si_4C_2 + C$ | 6.797 |
| $Si_4C_4$ | $Si + Si_3C_4$ | 4.723 |
| | $Si_2 + Si_2C_4$ | 5.172 |
| | $Si_3 + SiC_4$ | 5.251 |
| | $Si_4 + C_4$ | 7.272 |
| | $Si_4C + C_3$ | 6.044 |
| | $Si_4C_2 + C_2$ | 6.590 |
| | $Si_4C_3+C$ | 7.538 |

**Table 20.** Average coordination numbers for the ground state $Si_mC_n$ clusters.

| Figure | Silicon – Carbon ratio | Average coordination number |
|---|---|---|
| 1.1.1 | 1:1 | 1.0 |
| 1.2.1 | 1:2 | 2.0 |
| 1.3.1 | 1:3 | 1.5 |
| 1.4.1 | 1:4 | 1.6 |
| 2.1.1 | 2:1 | 2.0 |
| 2.2.1 | 2:2 | 2.5 |
| 2.3.1 | 2:3 | 1.6 |
| 2.4.1 | 2:4 | 1.6 |
| 3.1.1 | 3:1 | 2.5 |
| 3.2.1 | 3:2 | 2.8 |
| 3.3.1 | 3:3 | 3.3 |
| 3.4.1 | 3:4 | 2.6 |
| 4.1.1 | 4:1 | 3.2 |
| 4.2.1 | 4:2 | 2.7 |
| 4.3.1 | 4:3 | 2.9 |
| 4.4.1 | 4:4 | 3.5 |

**Table 21.** Harmonic vibrational frequencies (in $cm^{-1}$) for the ground state $Si_mC_n$ clusters. Results are compared with the theoretical frequencies of Bertolus *et al*. [25, 60] for identical structures and with experimental data.

| Cluster | This work | Reference [25] | Experimental |
|---|---|---|---|
| SiC $C_v$ | 986.1 | | 962 |
| $SiC_2$ $C_{2v}$ | 299.9, 806.8, 1825.5 | 216.8, 811.9, 1823.6 | 160.4, 824.3, 1741.3 |
| $SiC_3$ $C_{2v}$ | 138.7, 138.7, 350.6, 350.6, 616.4, 1325.0, 1989.0 | | |
| $SiC_4$ $C_{\infty v}$ | 90.8, 90.9, 222.0, 222.2, 555.1, 555.4, 572.4, 1172.0, 1869.4, 2187.0 | 105.2, 256.8, 574.3,629.9, 1176.1, 1866.3, 2198.7 | 2095.5 |
| $Si_2C$ $C_{2v}$ | 51.1, 730.2, 1312.4 | 95.7, 785.9, 1257.0 | 839.5, 1188.4 |
| $Si_2C_2$ $D_{2h}$ | 187.2, 372.9, 512.9, 968.3, 980.6, 1161.4 | 200.0, 368.9, 513.1, 967.9, 974.6, 1164.2 | 382.2, 982.9 |
| $Si_2C_3$ $D_{\infty h}$ | 79.4, 79.4, 203.5, 203.5, 464.8, 548.0, 548.0, 907.3, 1575.9, 2062.7 | 90.2, 225.2, 465.2, 604.8 905.6, 1576.5, 2077.3 | 898.9, 1955.2 |
| $Si_2C_4$ $D_{\infty h}$ | 66.9, 66.9, 177.3, 177.3, 379.0, 379.0, 411.0, 647.0, 647.0, 732.6, 1251.4, 1861.6, 2063.4 | 71.8, 184.5, 390.0, 412.5, 625.3, 732.8, 1255.7, 1869.1, 2078.4 | 719.1, 1807.4 |
| $Si_3C$ $C_{2v}$ | 168.9, 303.1, 356.8, 510.8, 672.9, 1114.8 | 186.3, 306.9, 361.1, 511.7, 671.4, 1111.9 | 309.5, 357.6, 511.8, 658.2, 1101.4 |
| $Si_3C_2$ $C_{2v}$ | 138.6, 183.7, 186.7, 441.2, 446.6, 610.3, 714.6, 960.9, 1514.2 | 148.9, 189.5, 195.7, 456.6, 467.5, 607.0, 710.7, 956.5, 1525.8 | 597.8, 681.1, 956.7 |
| $Si_3C_3$ $C_s$ | 211.1, 270.6, 299.4, 470.0, 507.0, 525.7, 584.8, 686.5, 732.1, 935.4, 1577.7 | 196.2, 210.8, 267.3, 278.7, 439.2, 471.4, 480.1, 554.8, 657.5, 665.7, 912.2, 1624.8 | 719.1, 1807.4 |
| $Si_3C_4$ | 128.8, 196.3, 230.6, 343.2, 354.0, 366.3, 377.8, 429.5, 479.4, 541.9, 584.7, 831.3, 860.1, 1685.7, 1731.4 | 138.2, 194.9, 236.1, 345.4, 353.9, 385.3, 385.5, 460.4, 479.3, 550.3, 589.8, 831.9, 869.9, 1682.6, 1741.5 | |
| $Si_4C$ $C_{3v}$ | 236.3, 324.2, 339.0, 417.6, 689.4, 741.84 | 238.0, 322.4, 348.6, 424.2, 692.7, 739.5 | |
| $Si_4C_2$ $C_{2v}$ | 215.6, 269.3, 318.8, 385.1, 423.5, 438.3, 504.0, 523.3, 746.8, 881.8, 991.7 | | |
| $Si_4C_3$ $C_1$ | 123.8, 177.8, 203.9, 257.5, 284.5, 337.1, 412.7, 480.7, 503.9, 528.2, 569.5, 583.0, 890.8, 1044.6, 1588.5 | 130.9, 185.2, 197.8, 260.1, 288.5, 353.8, 418.6, 507.5, 507.8, 531.7, 574.2, 594.7, 892.4, 1049.5, 1598.7 | |
| $Si_4C_4$ $C_{2v}$ | 120.1, 173.9, 228.61, 259.3, 281.0, 334.7, 354.7, 394.9, 397.6, 417.0, 486.8, 537.4, 558.8, 682.8, 805.1, 840.8, 1267.3, 1803.0 | 119.6, 178.6, 232.1, 267.9, 293.0, 335.7, 353.8, 400.5, 412.4, 424.9, 497.6, 540.1, 566.5, 686.6, 807.8, 840.6, 1273.5, 1816.4 | |

**Table 22.** Mulliken charge distributions for the ground state $Si_mC_n$ clusters.

| Cluster | Charge distribution | | | |
|---|---|---|---|---|
| SiC | Si 0.280 | | C -0.280 | |
| $SiC_2$ | Si 0.389 | | C -0.194 | C -0.194 |
| $SiC_3$ | Si 0.195 | | C -3.294 | C 3.129 |
| $SiC_4$ | Si 0.193 | | C -3.212<br>C 3.649 | C 0.285<br>C -0.916 |
| $Si_2C$ | Si 0.263 | Si 0.263 | C -0.527 | |
| $Si_2C_2$ | Si 0.251 | Si 0.251 | C -0.251 | C -0.251 |
| $Si_2C_3$ | Si 0.061 | Si 0.061 | C 2.494<br>C -1.308 | C -1.308 |
| $Si_2C_4$ | Si 0.042 | Si 0.042 | C 1.566<br>C 1.566 | C -1.608<br>C -1.608 |
| $Si_3C$ | Si 0.109<br>Si 0.109 | Si 0.017 | C -0.235 | |
| $Si_3C_2$ | Si 0.220<br>Si 0.017 | Si 0.017 | C -0.127<br>C -0.127 | |
| $Si_3C_3$ | Si 0.236<br>Si 0.235 | Si 0.068 | C -0.200<br>C -0.137 | C -0.202 |
| $Si_3C_4$ | Si 0.357<br>Si 0.109 | Si 0.241 | C -0.831<br>C 0.574 | C 0.203<br>C -0.656 |
| $Si_4C$ | Si 0.119<br>Si 0.119 | Si 0.119<br>Si -0.041 | C -0.317 | |
| $Si_4C_2$ | Si 0.200<br>Si 0.200 | Si 0.200<br>Si -0.103 | C -0.147<br>C -0.350 | |
| $Si_4C_3$ | Si 0.256<br>Si 0.108 | Si 0.057<br>Si -0.098 | C -0.158<br>C 0.100 | C -0.266 |
| $Si_4C_4$ | Si 0.281<br>Si 0.281 | Si -0.016<br>Si 0.331 | C -0.130<br>C -0.308 | C -0.130<br>C -0.308 |

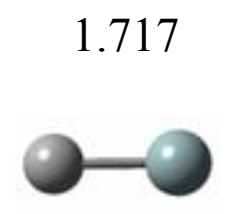

**1.1.1**

**Figure1**. SiC neutral cluster

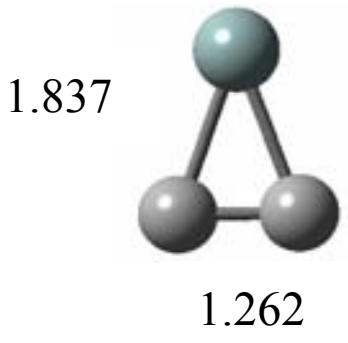

**1.2.1**

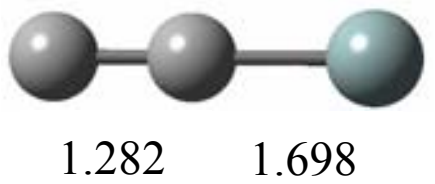

**1.2.2**

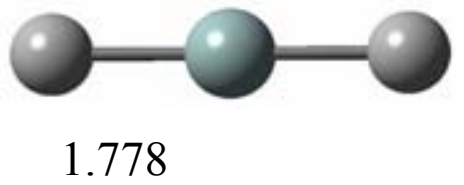

**1.2.3**

**Figure 2**. $SiC_2$ neutral clusters

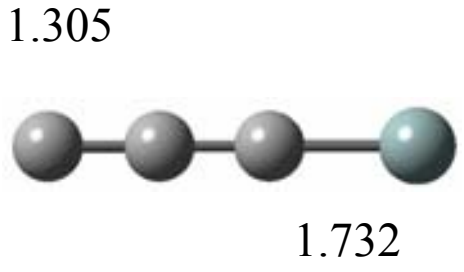

**1.3.1**

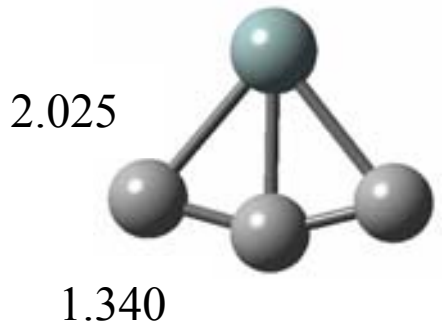

**1.3.2**

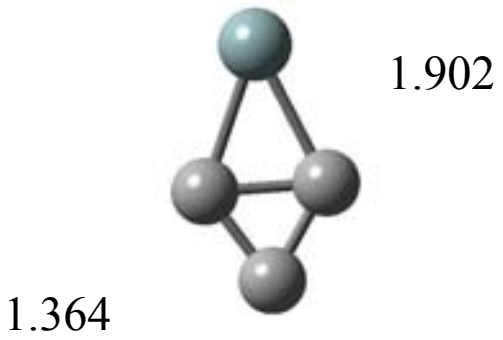

**1.3.3**

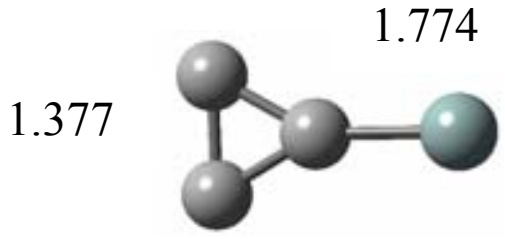

**1.3.4**

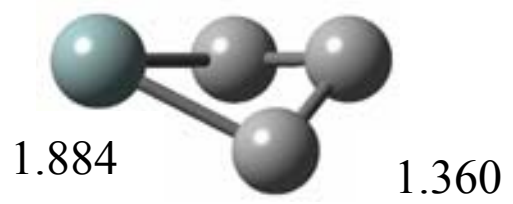

**1.3.5**

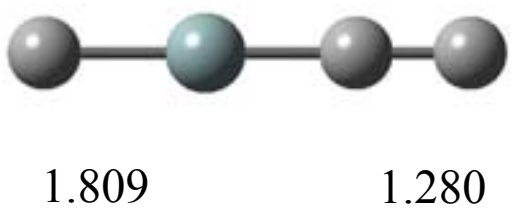

**1.3.6**

**Figure 3**. $SiC_3$ neutral clusters

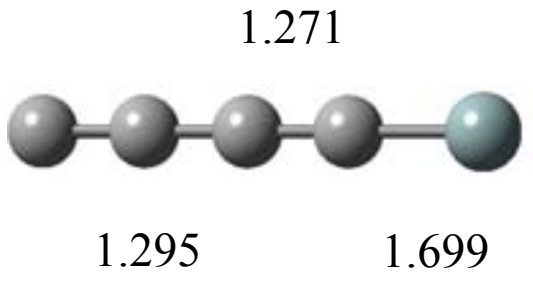

**1.4.1**

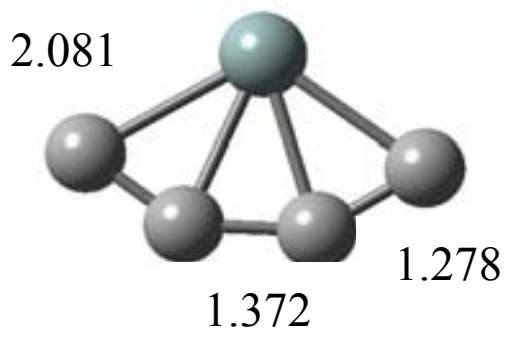

**1.4.2**

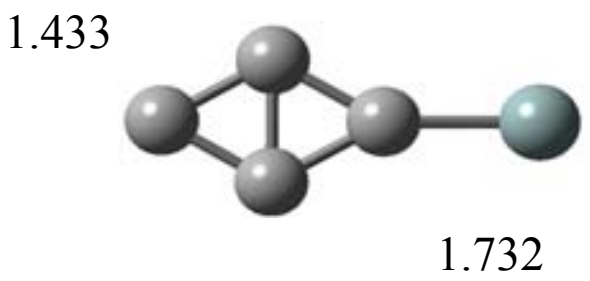

**1.4.3**

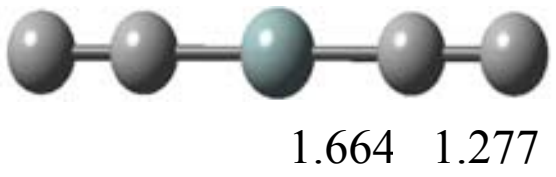

**1.4.4**

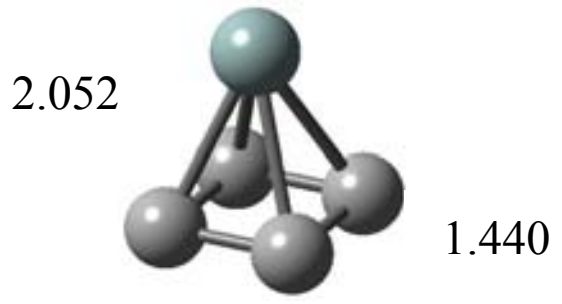

**1.4.5**

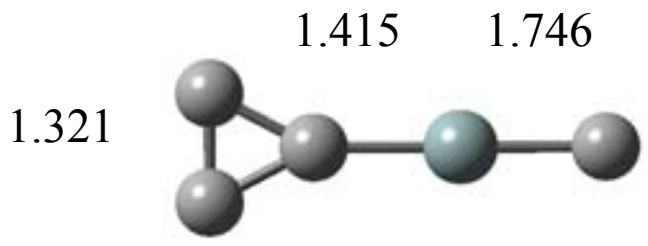

**1.4.6**

**Figure 4**. $SiC_4$ neutral clusters

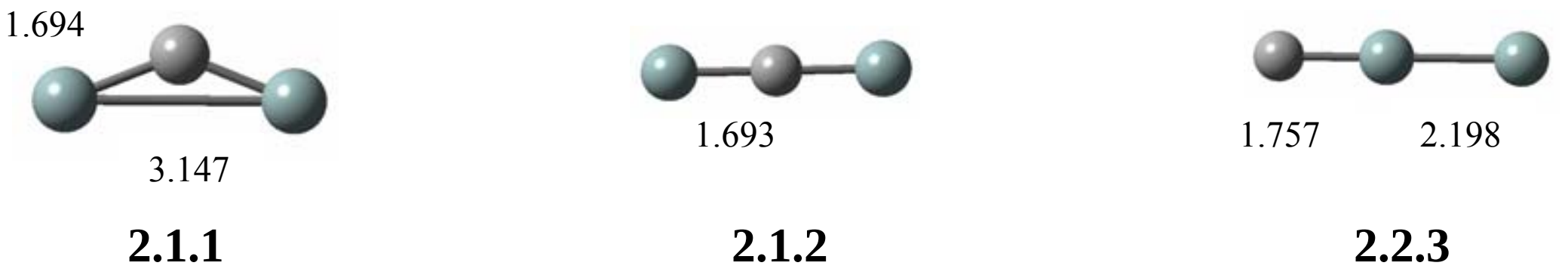

**Figure 5**. $Si_2C$ neutral clusters

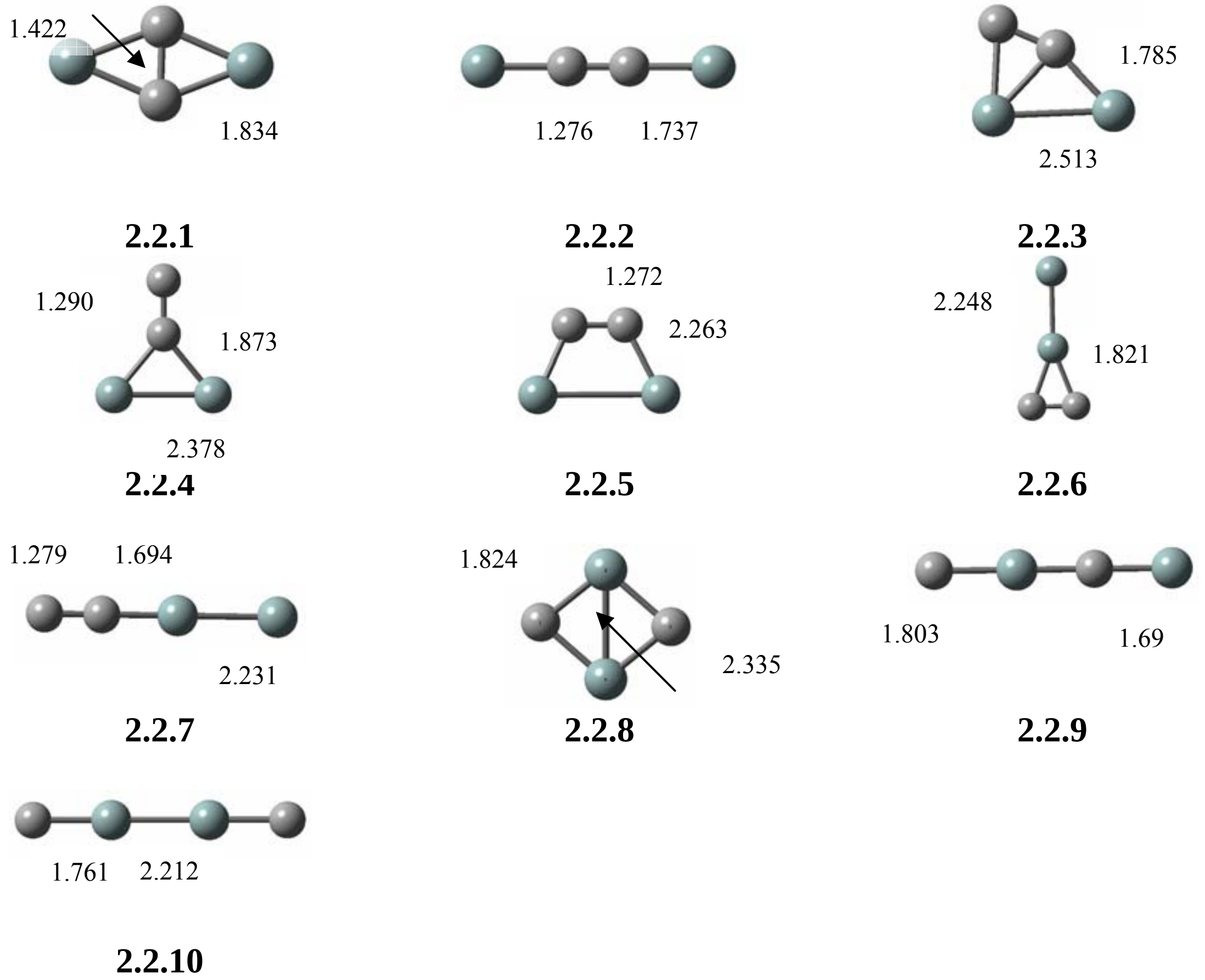

**Figure 6**. $Si_2C_2$ neutral clusters

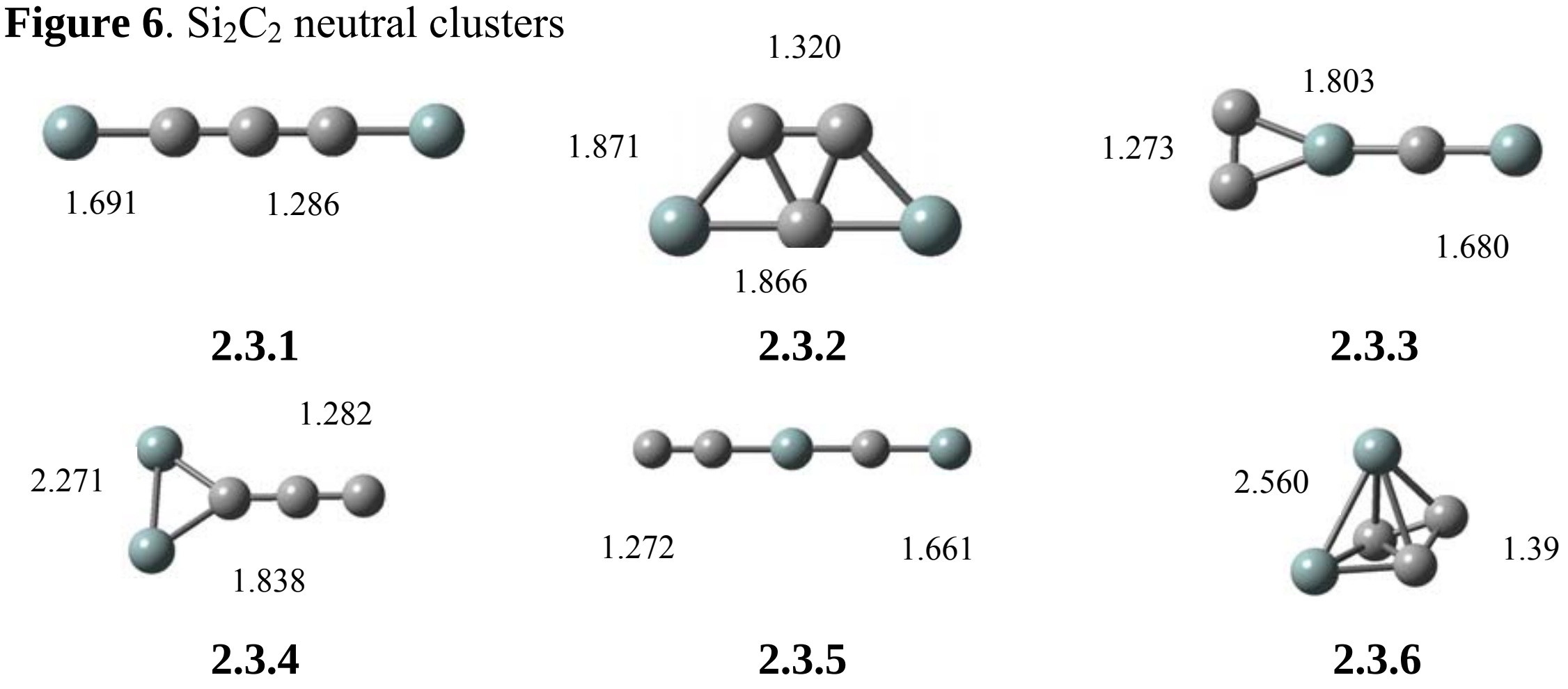

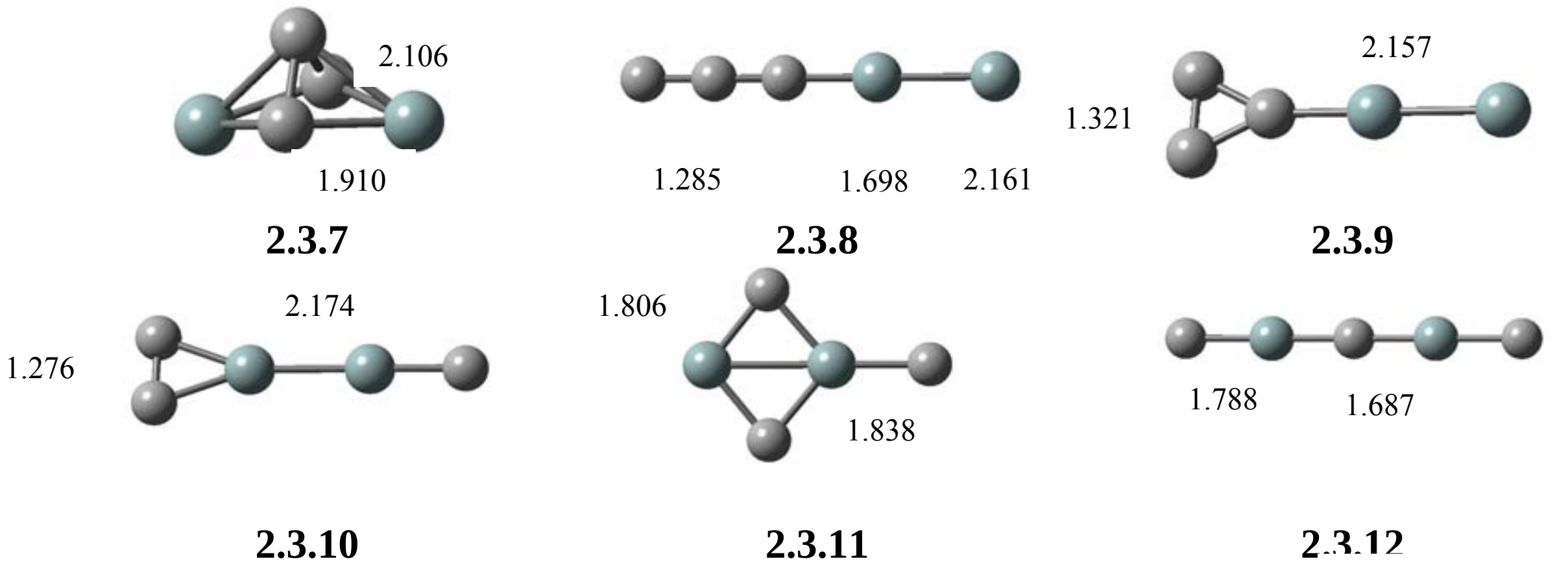

**Figure 7**. $Si_2C_3$ neutral clusters

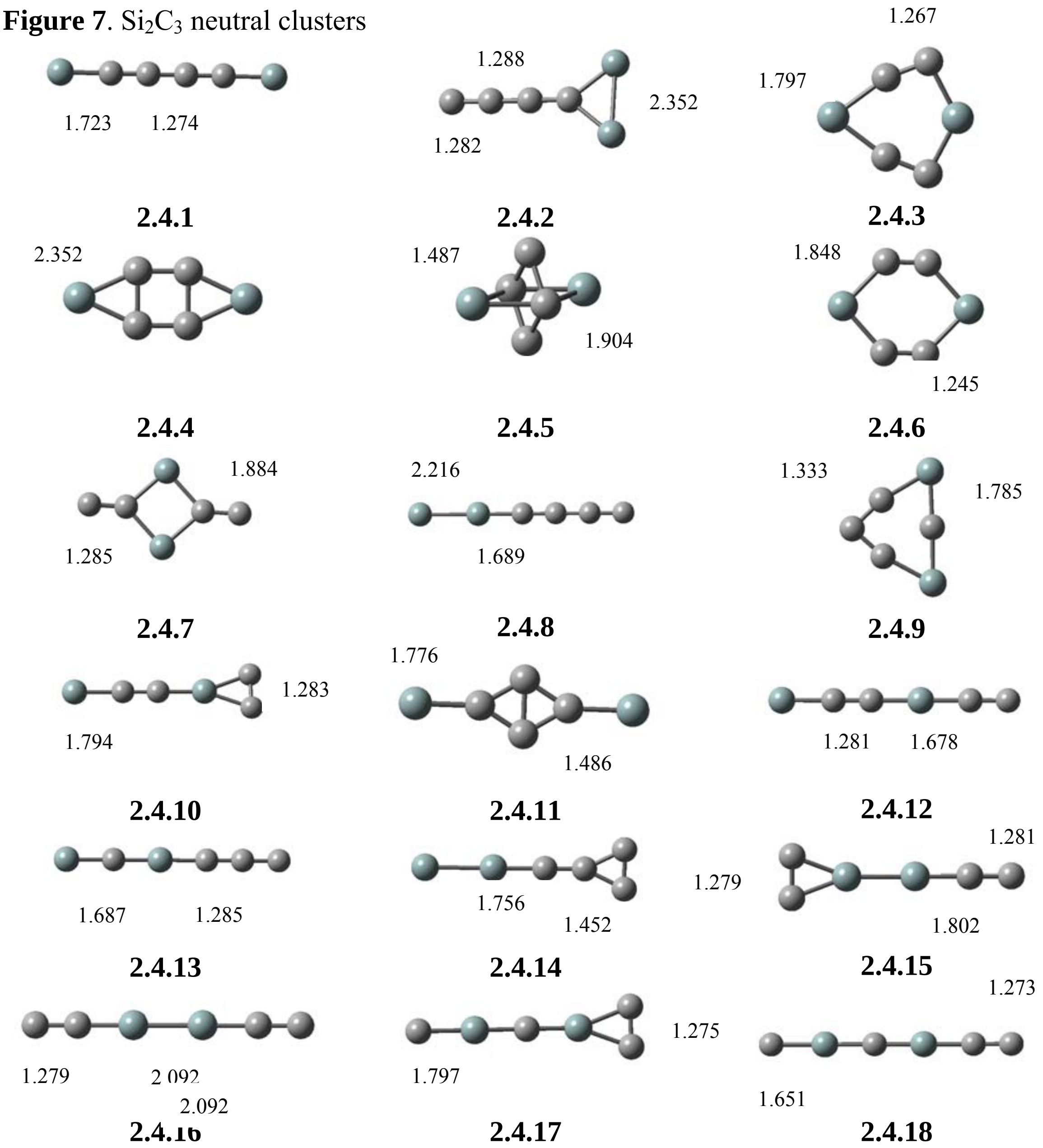

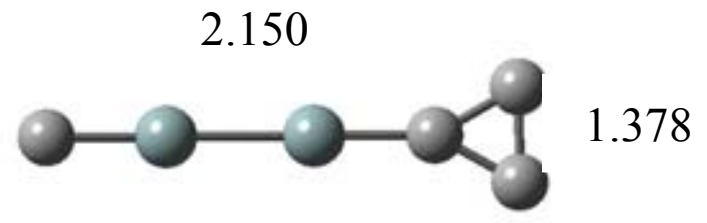

**2.4.19**

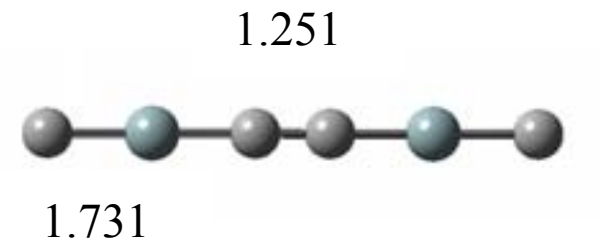

**2.4.20**

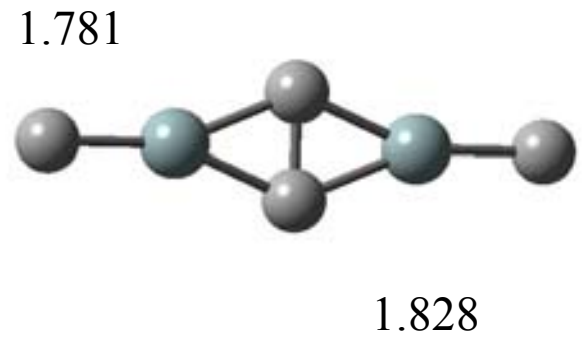

**2.4.21**

**Figure 8**. $Si_2C_4$ neutral clusters

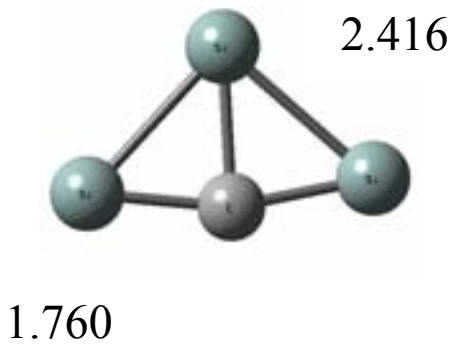

**3.1.1**

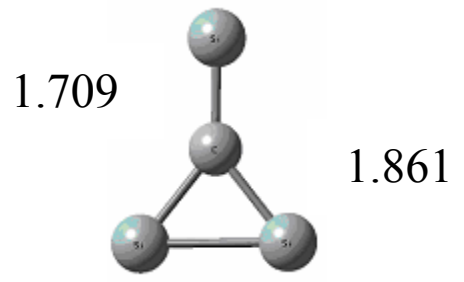

**3.1.2**

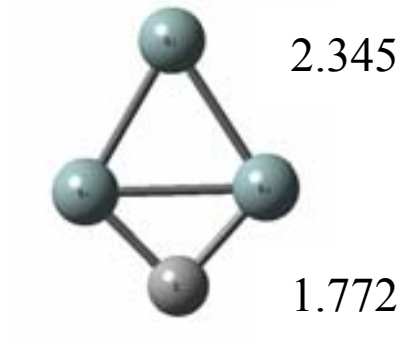

**3.1.3**

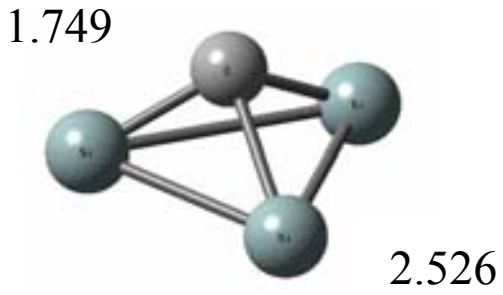

**3.1.4**

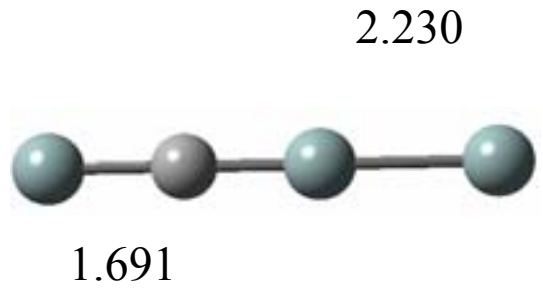

**3.1.5**

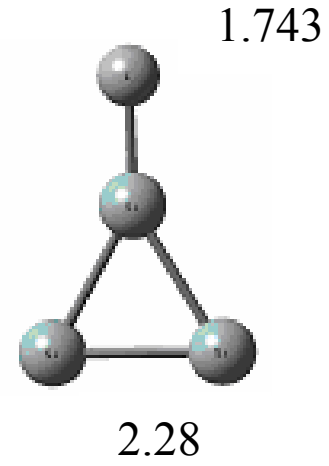

**3.1.6**

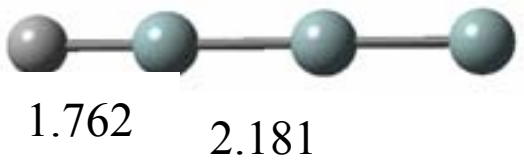

**3.1.7**

**Figure 9**. $Si_3C$ neutral clusters

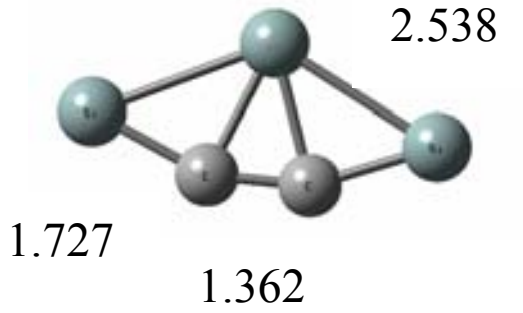

**3.2.1**

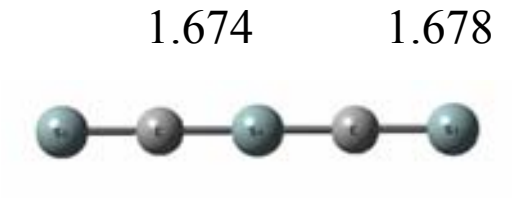

**3.2.2**

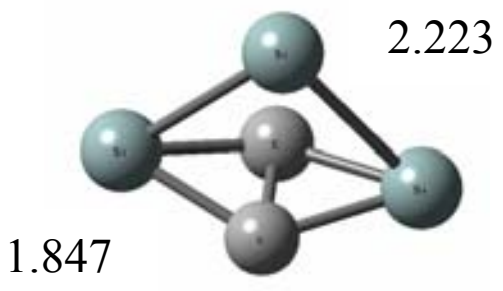

**3.2.3**

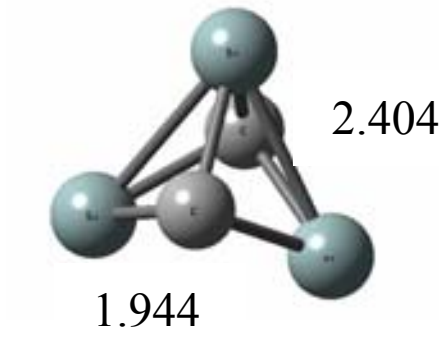

**3.2.4**

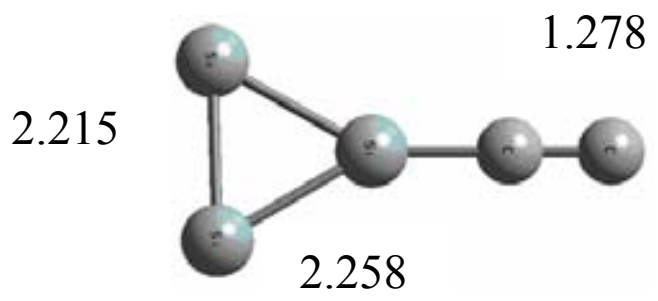

**3.2.5**

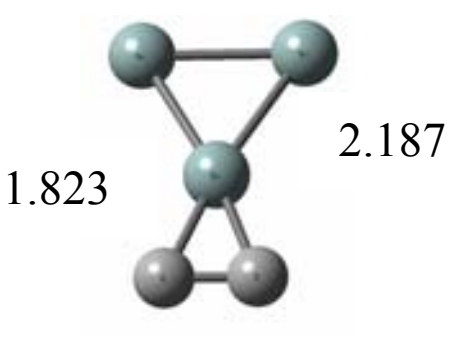

**3.2.6**

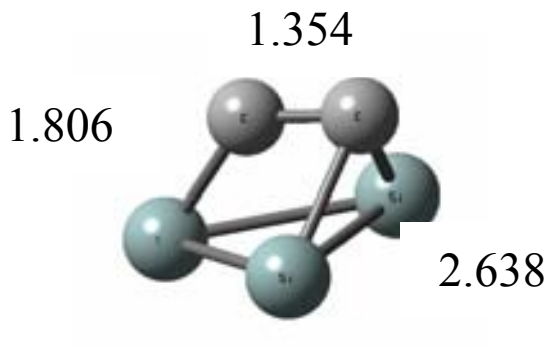

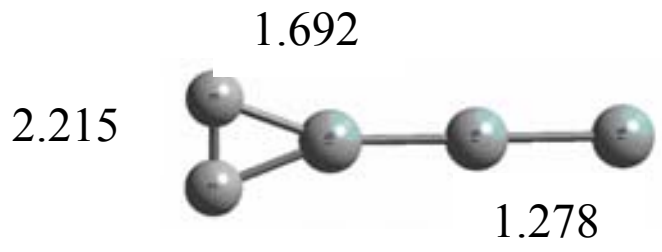

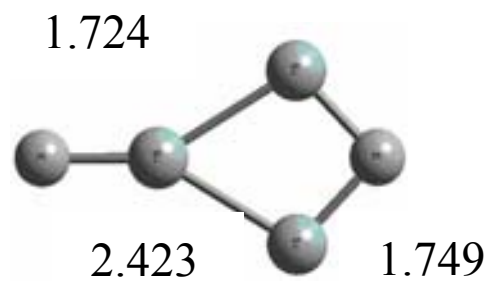

**3.2.7** **3.2.8** **3.2.9**

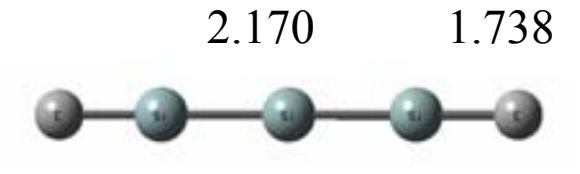

**3.2.10**

**Figure 10**. $Si_3C_2$ neutral clusters.

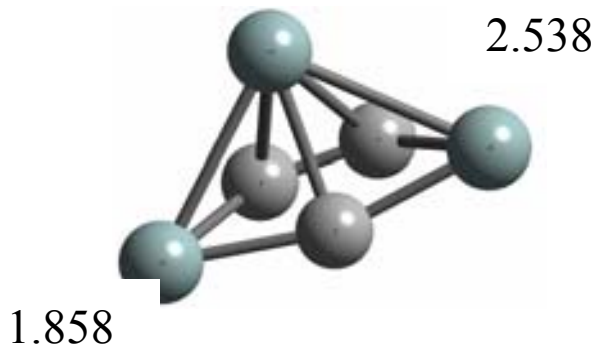

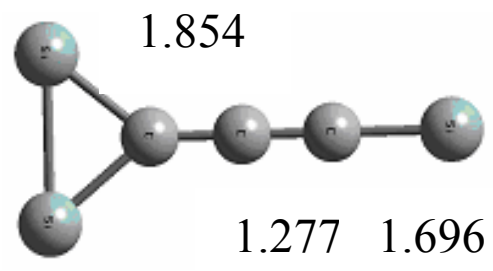

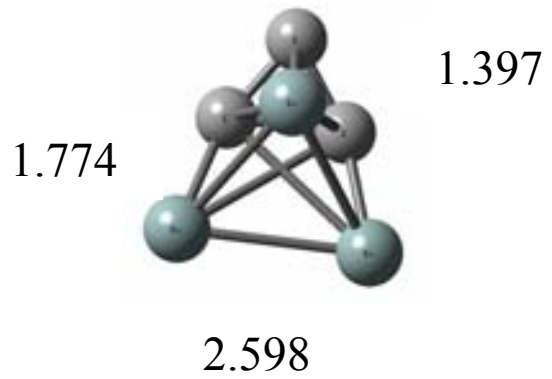

**3.3.1** **3.3.2** **3.3.3**

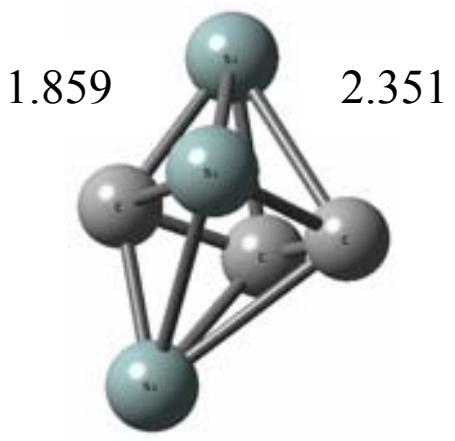

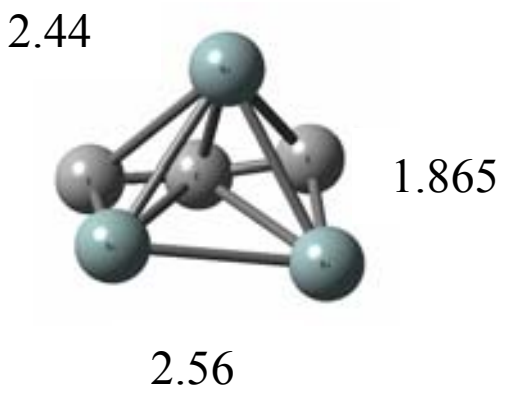

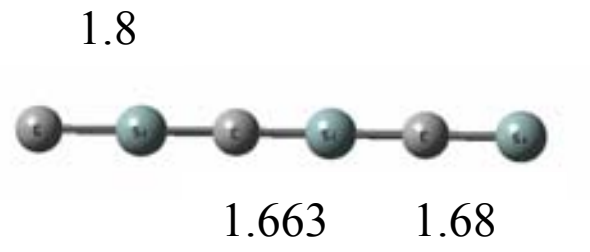

**3.3.4** **3.3.5** **3.3.6**

**Figure 11**. $Si_3C_3$ neutral clusters.

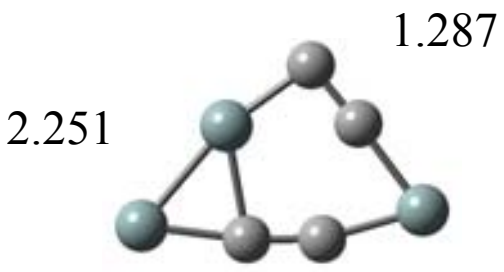

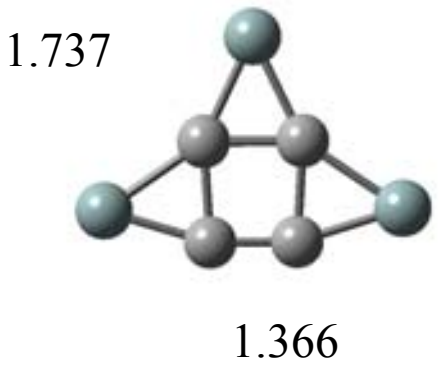

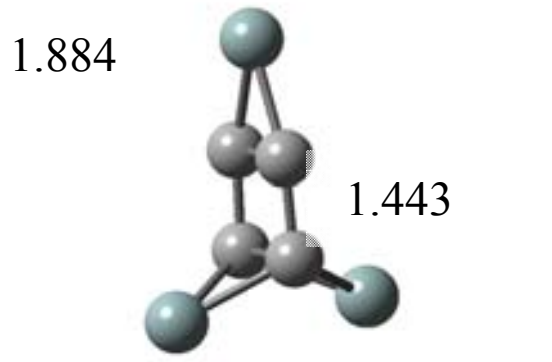

3.4.1 3.4.2 3.4.3

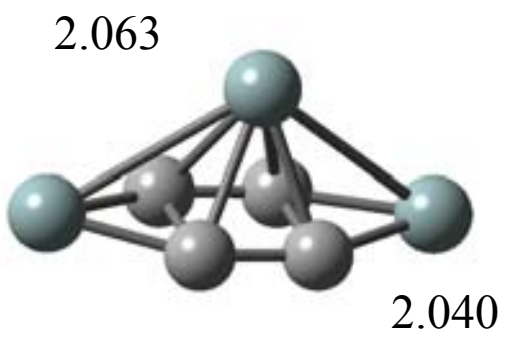

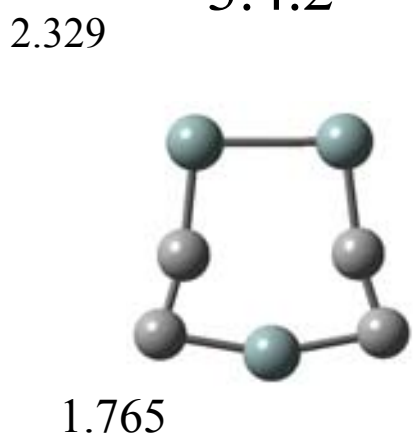

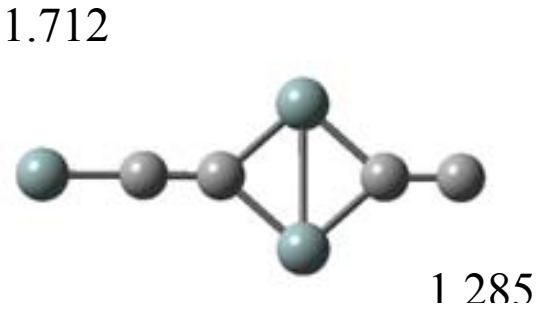

3.4.4 3.4.5 3.4.6

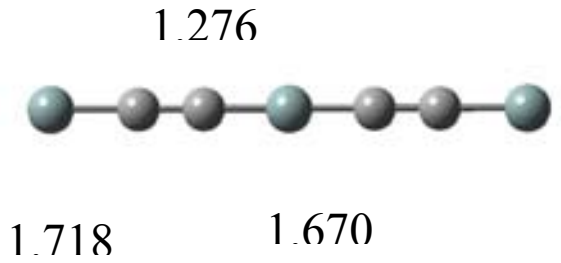

3.4.7

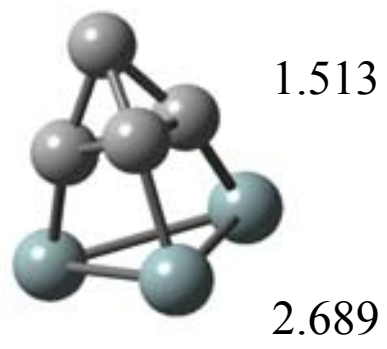

3.4.8

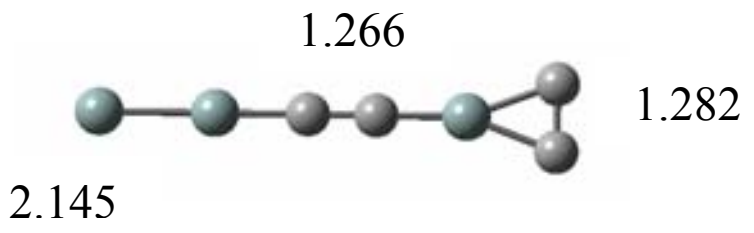

3.4.9

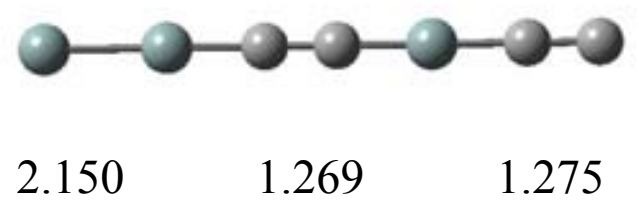

3.4.10

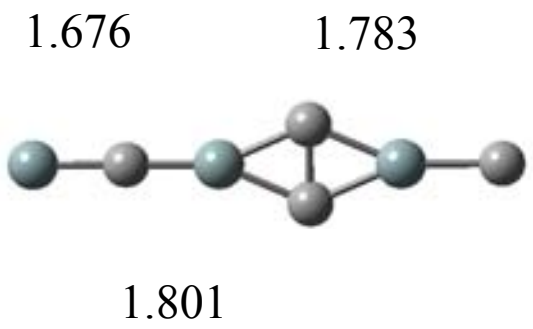

3.4.11

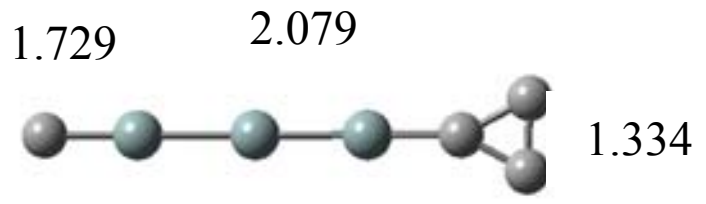

3.4.12

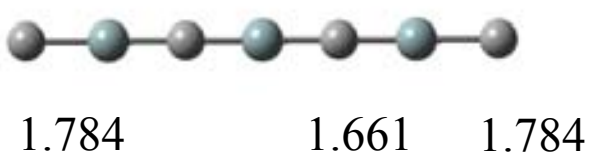

3.4.13

**Figure 12**. $Si_3C_4$ neutral clusters

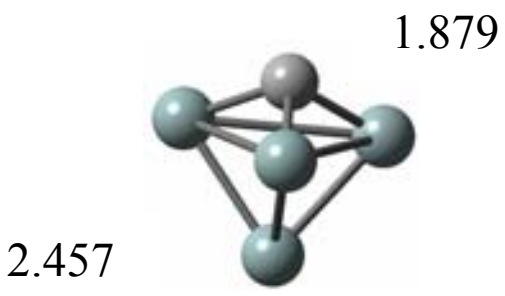

**4.1.1**

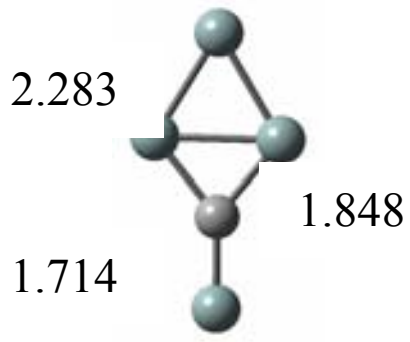

**4.1.2**

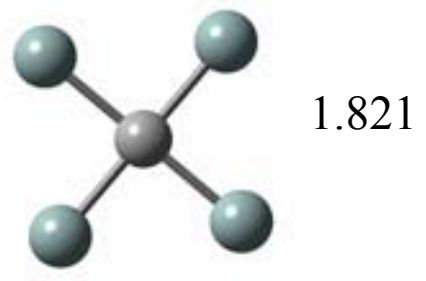

**4.1.3**

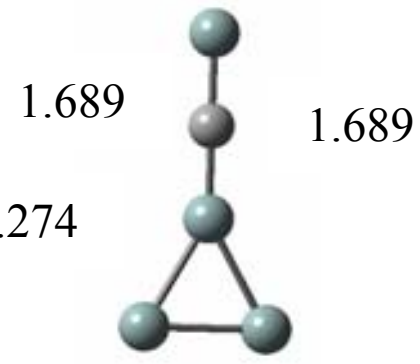

**4.1.4**

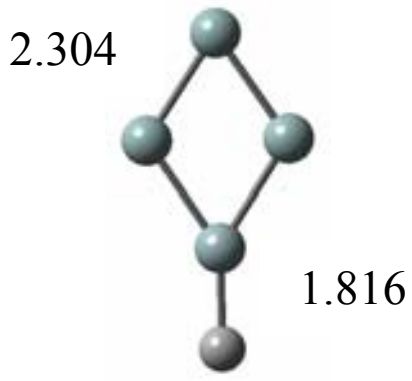

**4.1.5**

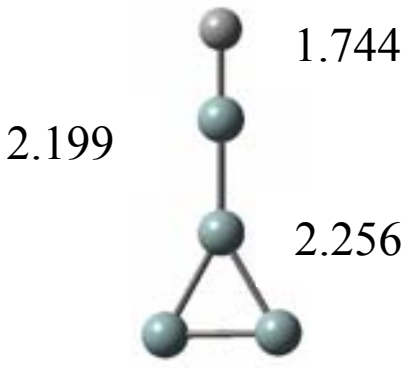

**4.1.6**

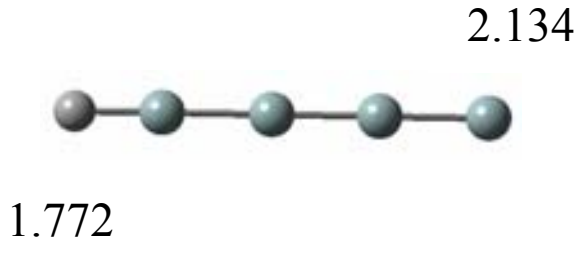

**4.1.7**

**Figure 13**. $Si_4C$ neutral clusters.

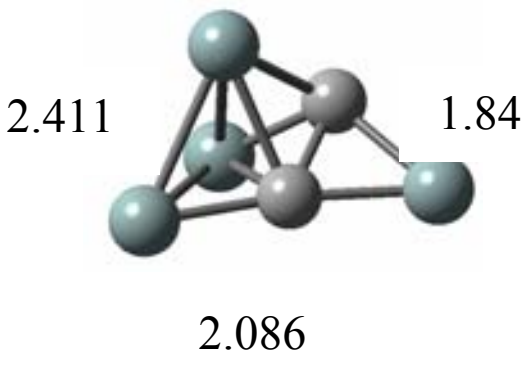

**4.2.1**

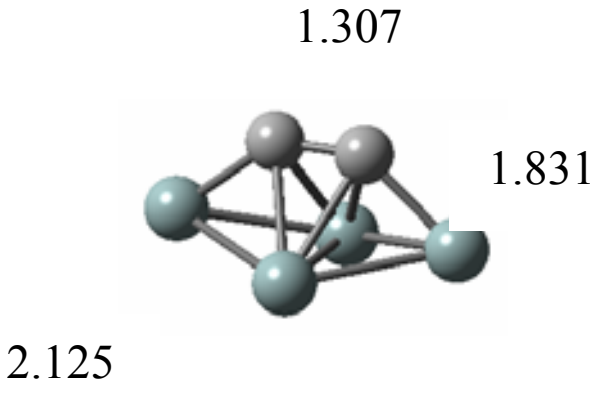

**4.2.2**

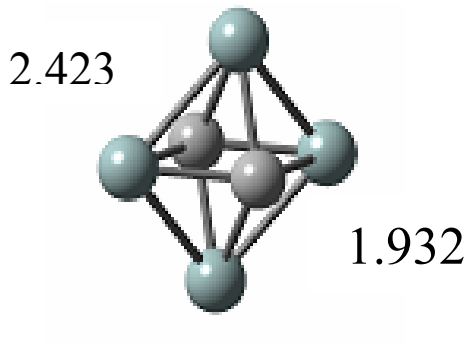

**4.2.3**

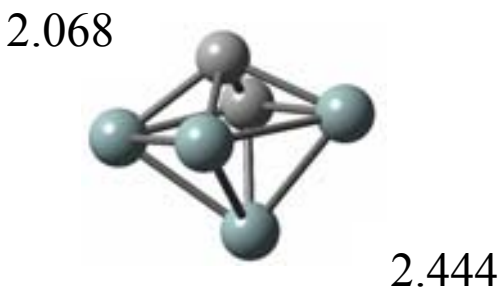

**4.2.4**

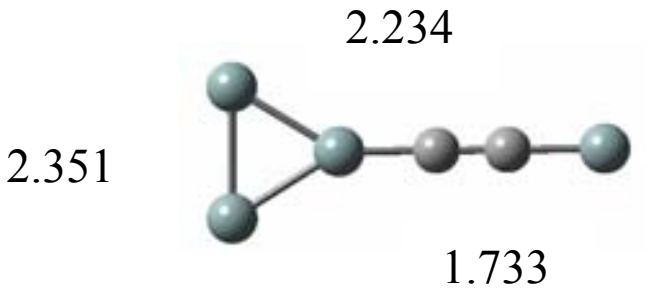

**4.2.5**

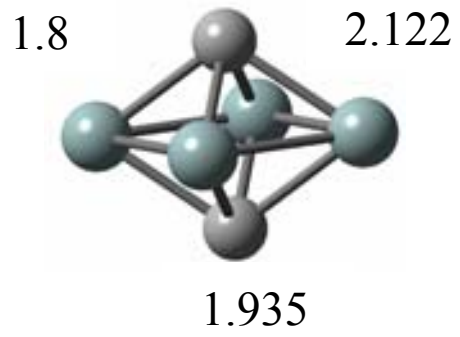

**4.2.6**

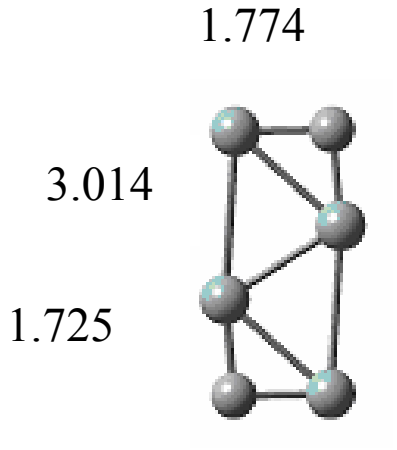

**4.2.7**

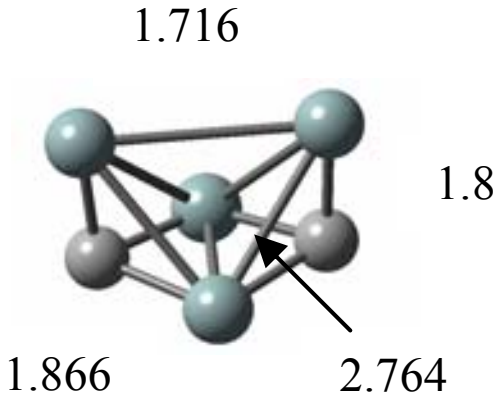

**4.2.7**

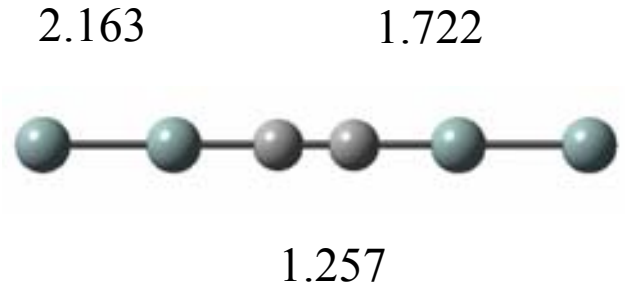

**4.2.9**

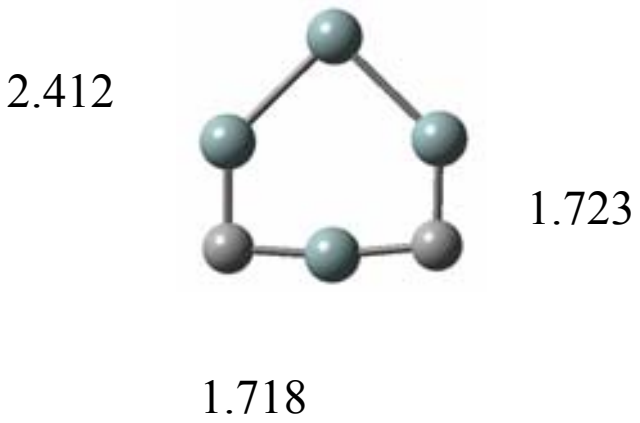

**4.2.10**

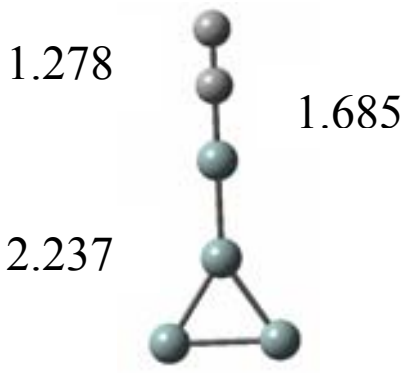

**4.2.11**

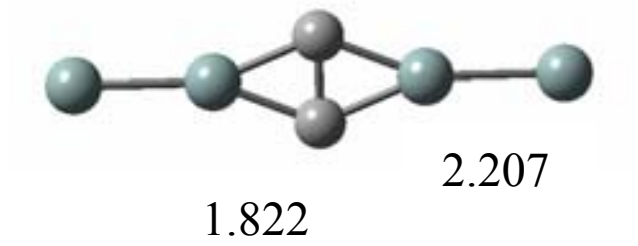

**4.2.12**

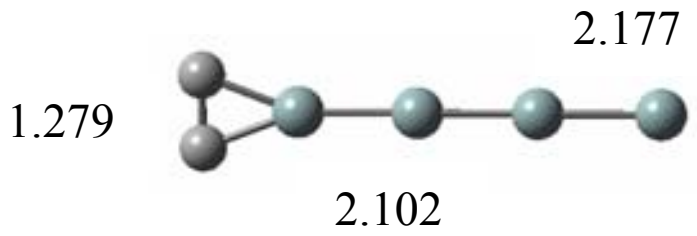

**4.2.13**

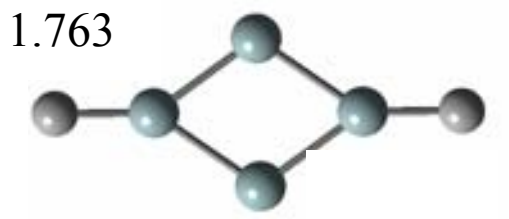

**4.2.14**

**Figure 14.** $Si_4C_2$ neutral clusters.

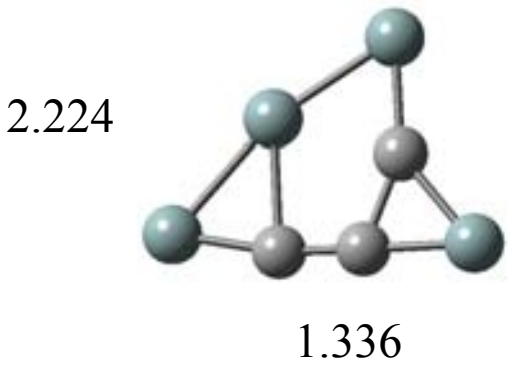

**4.3.1**

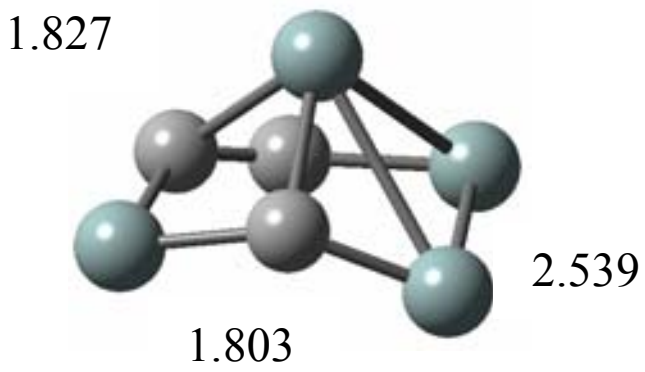

**4.3.2**

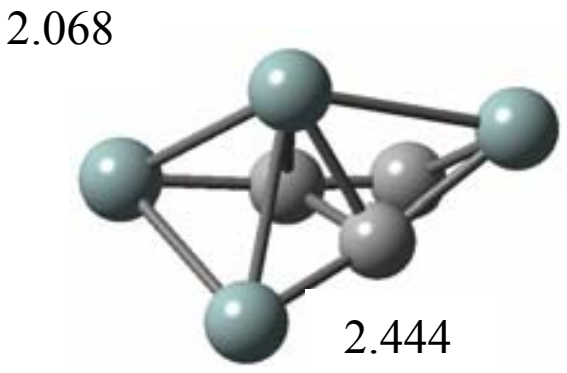

**4.3.3**

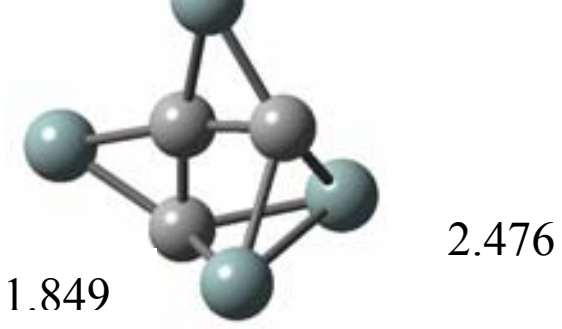

**4.3.4**

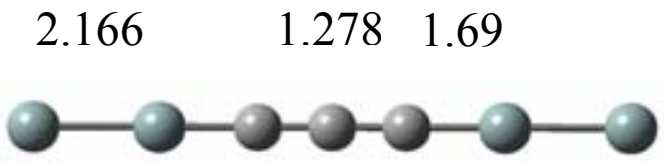

**4.3.5**

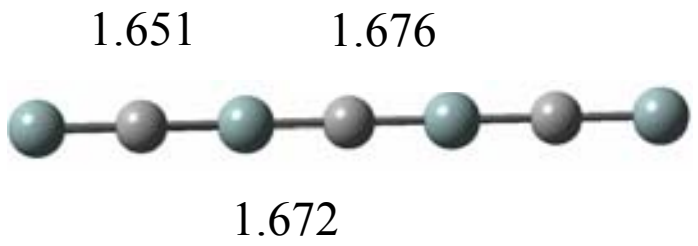

**4.3.6**

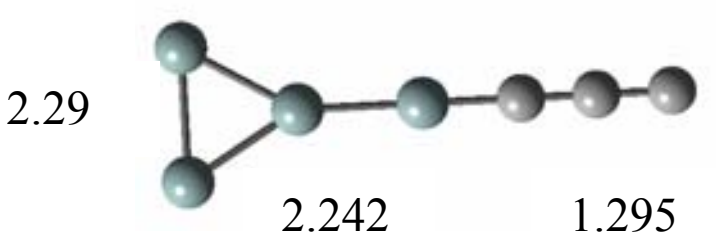

**4.3.7**

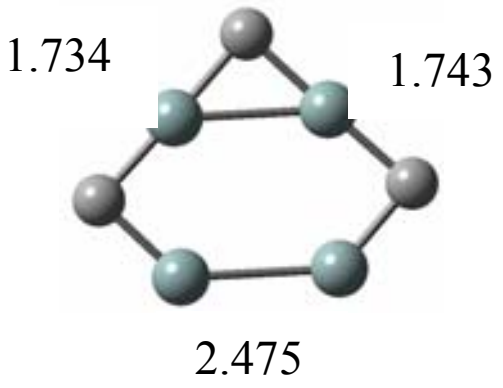

**4.3.8**

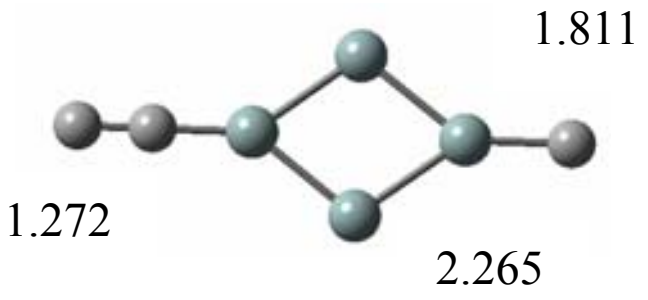

**4.3.9**

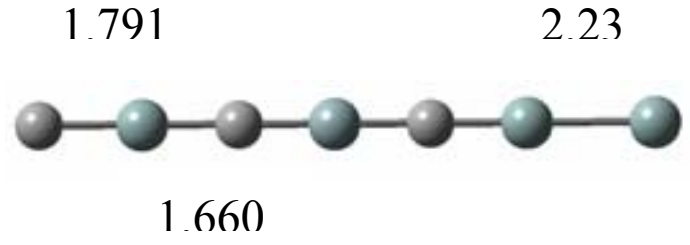

**4.3.10**

**Figure 15.** $Si_4C_3$ neutral clusters.

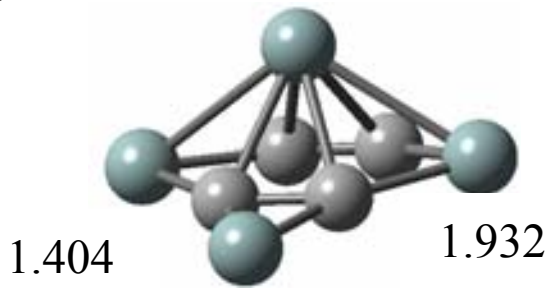

**4.4.1**

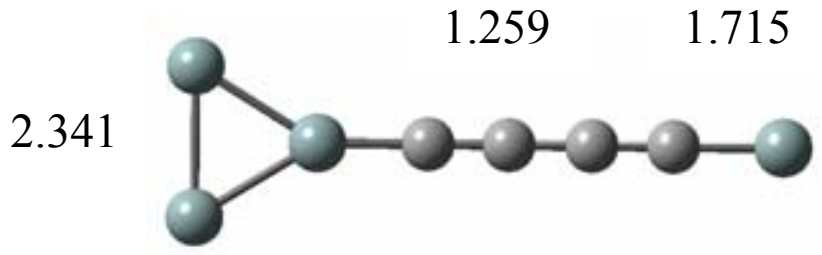

**4.4.2**

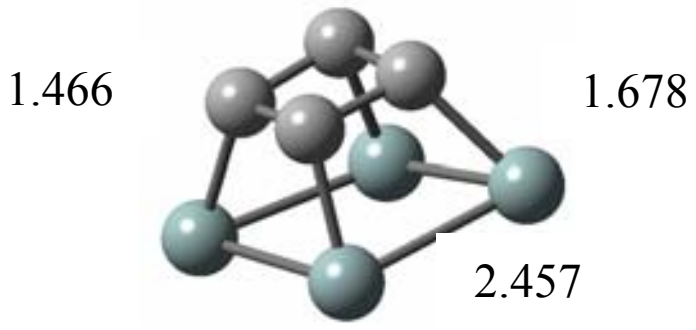

**4.4.3**

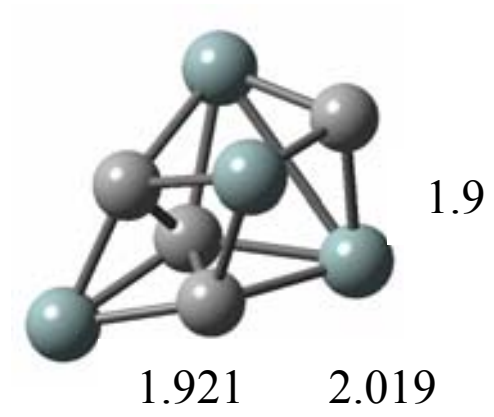

**4.4.4**

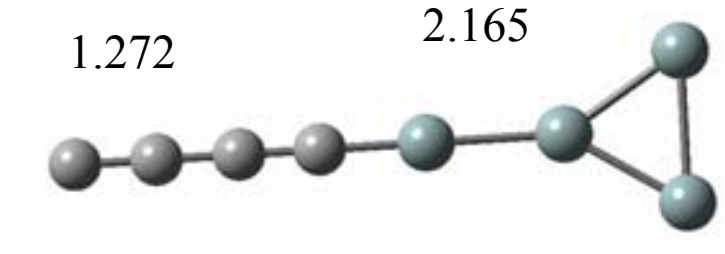

**4.4.5**

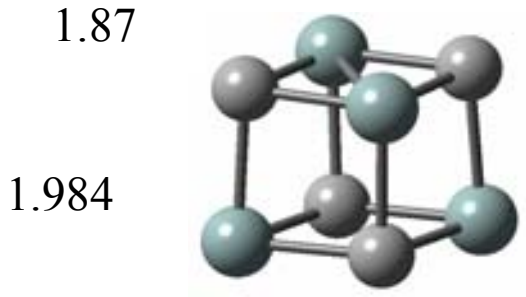

**4.4.6**

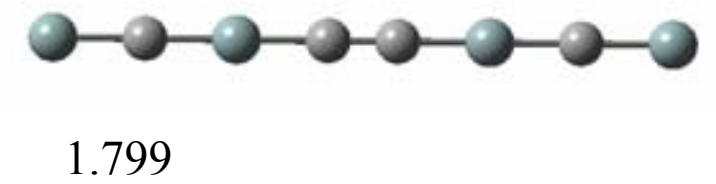

**4.4.7**

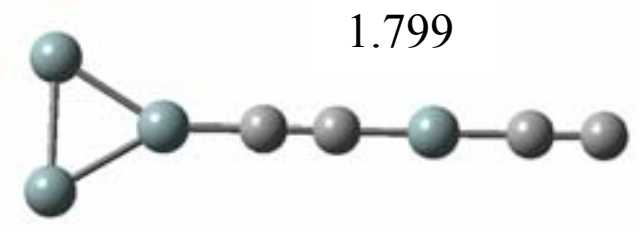

**4.4.8**

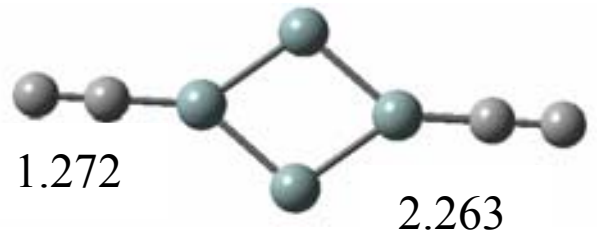

**4.9**

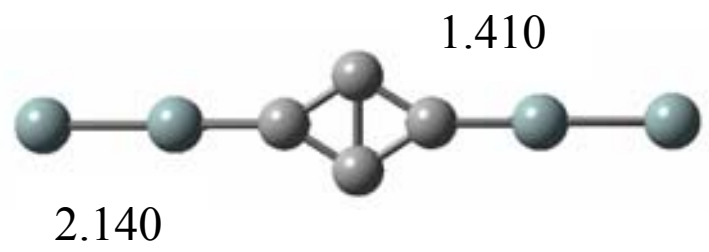

**4.4.10**

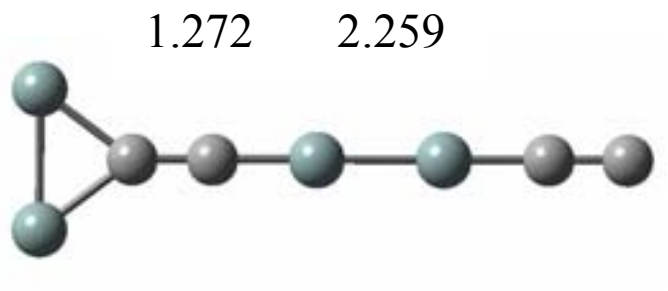

**4.4.11**

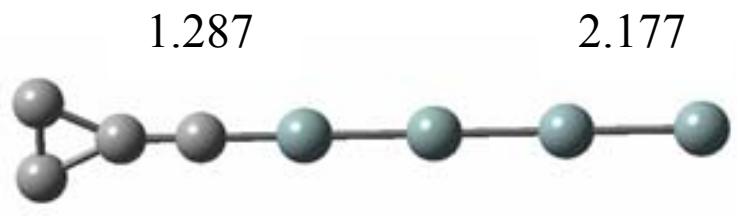

**4.4.12**

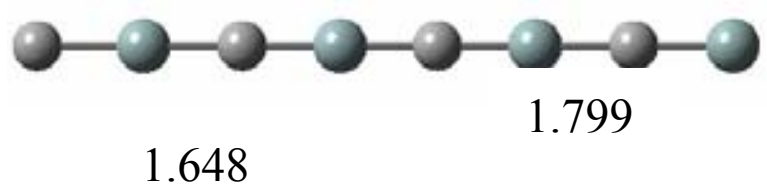

**4.4.13**

**Figure 16.** $Si_4C_4$ neutral clusters.

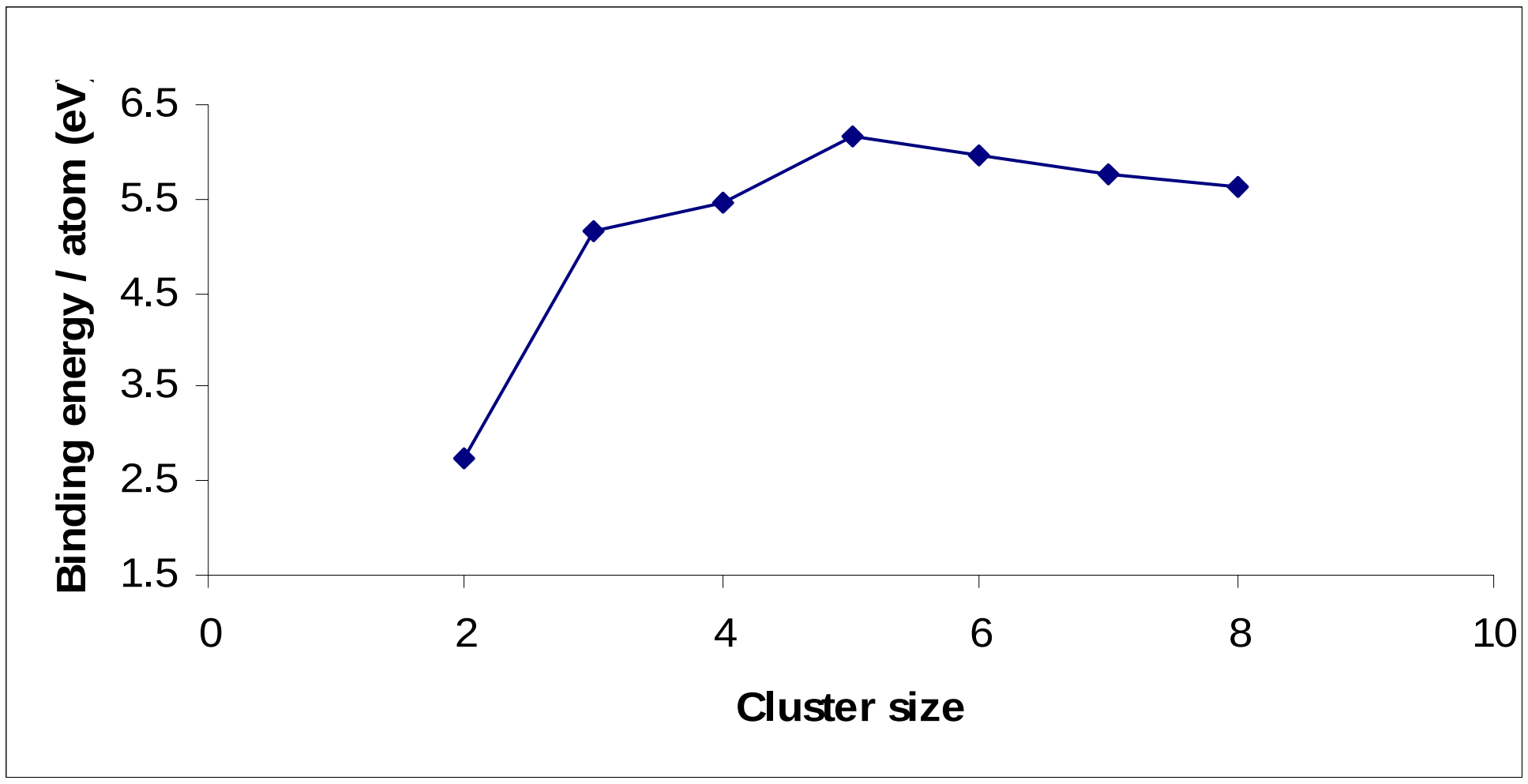

Figure 17. Binding energy per atom (in eV) versus the number of atoms in the cluster.

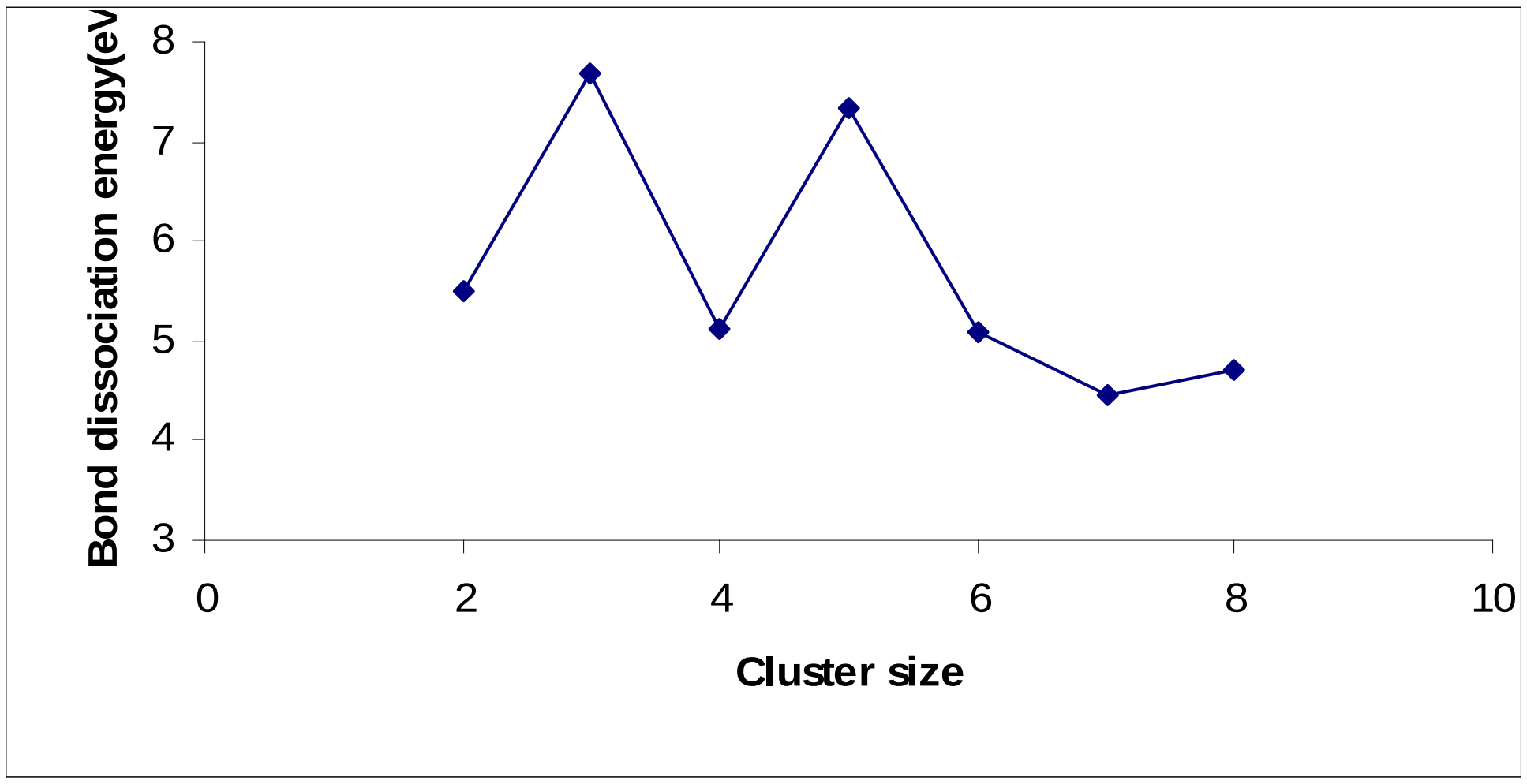

Figure 18. Bond dissociation energy (in eV) versus the number of atoms in the cluster.

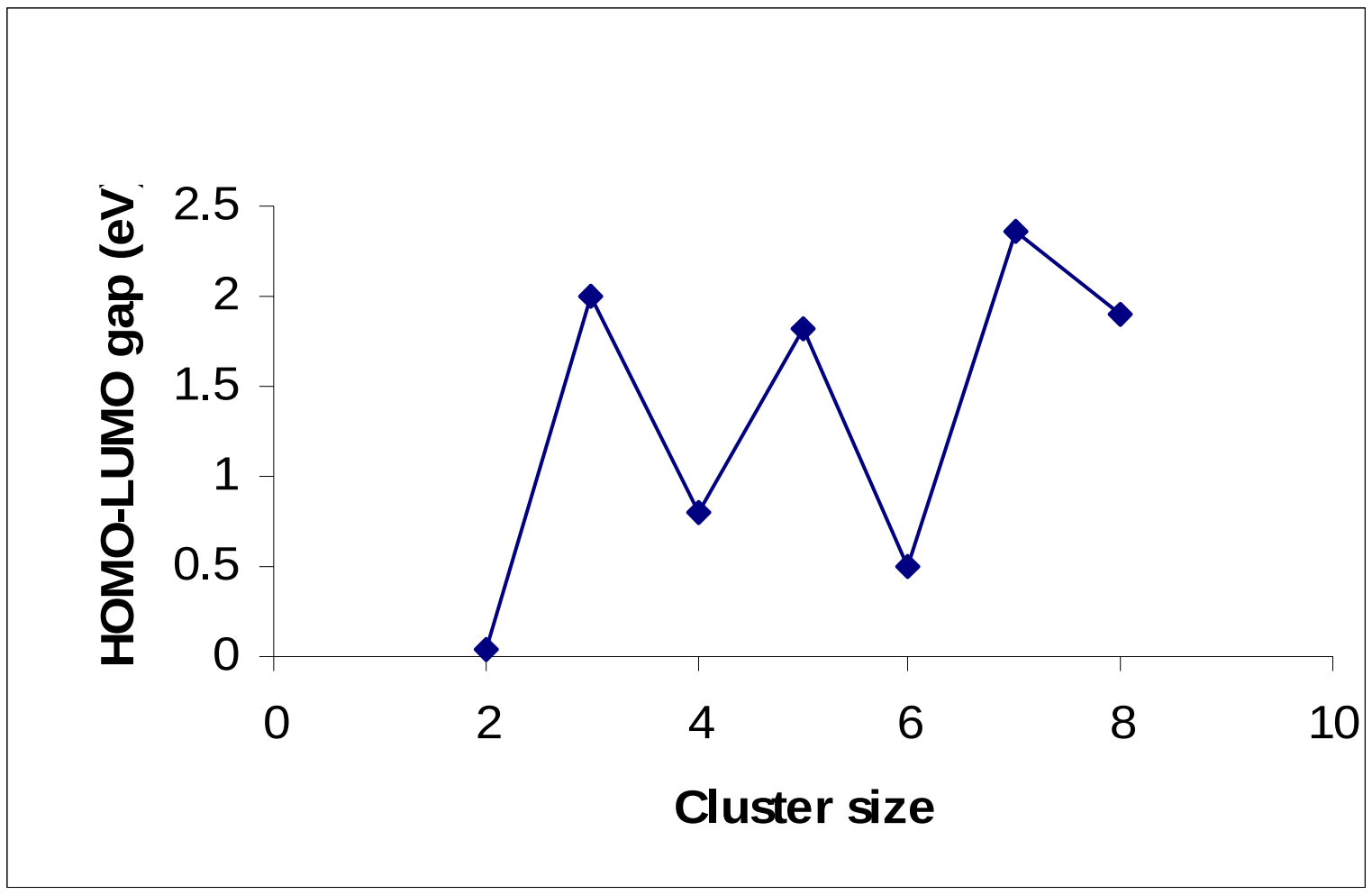

Figure 19. HOMO-LUMO gap (in eV) versus the number of atoms in the cluster.

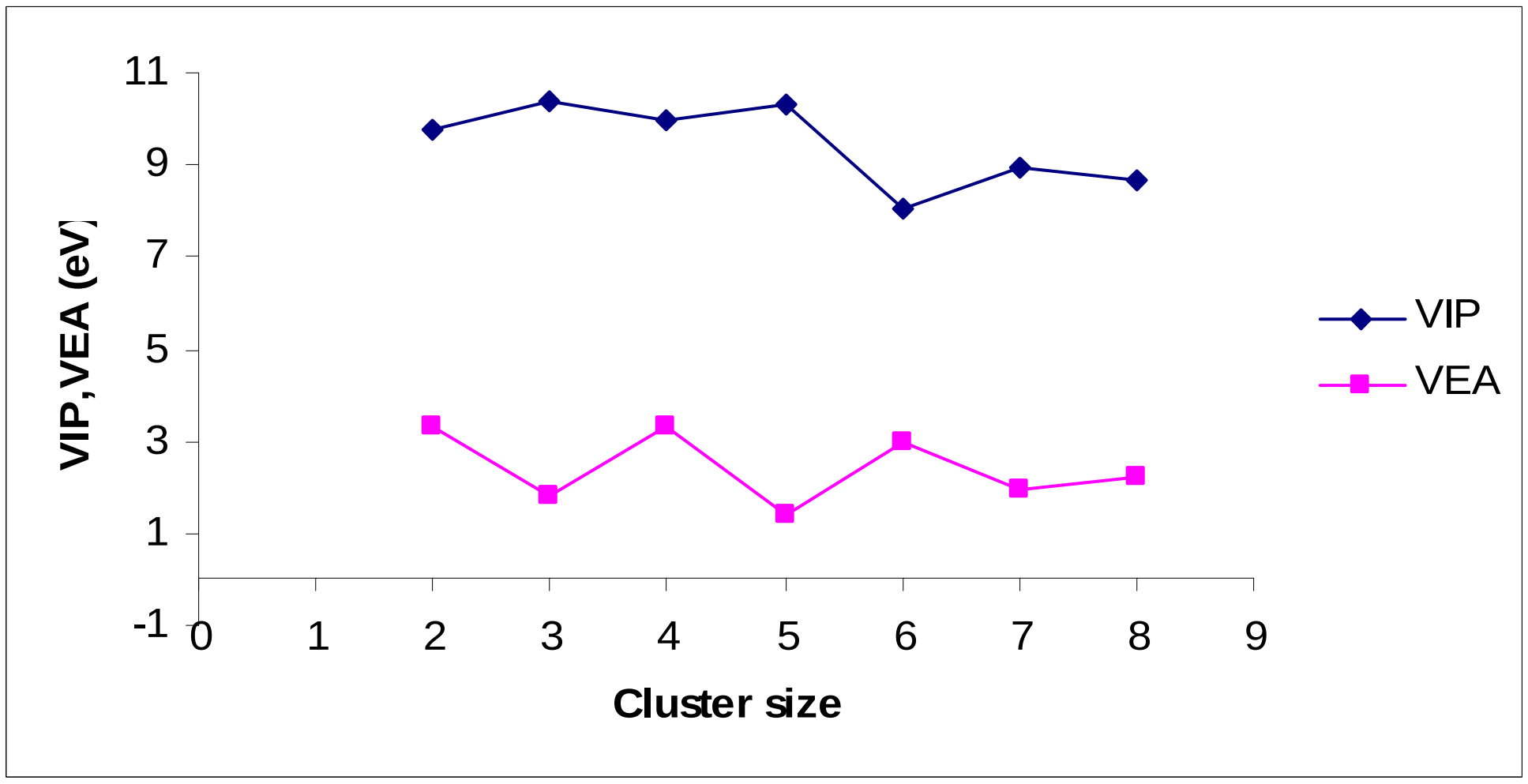

Figure 20. Vertical ionization potential and vertical electron affinity (in eV) versus the number of atoms in the cluster.